\documentclass[%
 reprint,
superscriptaddress,
amsmath,amssymb,
aps,
prb,
]{revtex4-2}

\usepackage{gensymb}
\usepackage{textcomp}
\usepackage{graphicx}
 \DeclareGraphicsExtensions{.eps,.bmp}
\usepackage{dcolumn}
\usepackage{bm}
\usepackage{hyperref}
\usepackage{multirow}

\begin{document}

\title{Thermodynamic insights into the intricate magnetic phase diagram of EuAl$_{4}$}

\author{William R. Meier}
\email{wmeier@utk.edu}%
\altaffiliation[Current Address: ]{Materials Science \& Engineering Department, University of Tennessee Knoxville, Knoxville, Tennessee 37996, USA}%
\affiliation{Materials Science \& Technology Division, Oak Ridge National Laboratory, Oak Ridge, Tennessee 37831, USA}%
 
\author{James R. Torres}
\affiliation{Materials Science \& Technology Division, Oak Ridge National Laboratory, Oak Ridge, Tennessee 37831, USA}%

\author{Raphael P. Hermann}
\affiliation{Materials Science \& Technology Division, Oak Ridge National Laboratory, Oak Ridge, Tennessee 37831, USA}%

\author{Jiyong Zhao}
\affiliation{Advanced Photon Source, Argonne National Laboratory 9700 S. Cass Avenue, Argonne, Illinois 60439, USA}%

\author{Barbara Lavina}
\affiliation{Advanced Photon Source, Argonne National Laboratory 9700 S. Cass Avenue, Argonne, Illinois 60439, USA}%
\affiliation{Center for Advanced Radiation Sources, The University of Chicago, Chicago, Illinois 60637, USA}%

\author{Brian C. Sales}
\affiliation{Materials Science \& Technology Division, Oak Ridge National Laboratory, Oak Ridge, Tennessee 37831, USA}%

\author{Andrew F. May}
\email{mayaf@ornl.gov}%
\affiliation{Materials Science \& Technology Division, Oak Ridge National Laboratory, Oak Ridge, Tennessee 37831, USA}%

\date{\today}

\begin{abstract}

	The tetragonal intermetallic compound EuAl$_{4}$ hosts an exciting variety of low temperature phases. In addition to a charge density wave below 140\,K, four ordered magnetic phases are observed below 15.4\,K. Recently, a skyrmion phase was proposed based on Hall effect measurements under a $c$-axis magnetic field. We present a detailed investigation of the phase transitions in EuAl$_{4}$ under $c$-axis magnetic field. Our dilatometry, heat capacity, DC magnetometry, AC magnetic susceptibility, and resonant ultrasound spectroscopy measurements reveal three magnetic phase transitions not previously reported. We discuss what our results reveal about the character of the magnetic phases. Our first key result is a detailed $\bm{H}$\,$\parallel$\,$[001]$ magnetic phase diagram mapping the seven phases we observe. Second, we identify a new high-field phase, phase VII, which directly corresponds to the region were skyrmions have been suggested. Our results provide guidance for future studies exploring the complex magnetic interactions and spin structures in EuAl$_{4}$.
	
\end{abstract}

This manuscript has been authored by UT-Battelle, LLC under Contract No. DE-AC05-00OR22725 with the U.S. Department of Energy. The United States Government retains and the publisher, by accepting the article for publication, acknowledges that the United States Government retains a non-exclusive, paid-up, irrevocable, world-wide license to publish or reproduce the published form of this manuscript, or allow others to do so, for United States Government purposes. The Department of Energy will provide public access to these results of federally sponsored research in accordance with the DOE Public Access Plan (http://energy.gov/downloads/doe-public-access-plan).

\maketitle

\section{Introduction}
\label{sec:Introduction}

Topological spin textures, such as skyrmions and merons, have generated significant interest for their promise for next generation electronic devices\cite{Tokura2020_MagneticSkyrmionMaterials,Goebel2021_BeyondSkyrmions,Back2020_SkyrmionRoadmap}. These phases are characterized by patterns of swirling magnetic moments built from superimposed, incommensurate magnetic modulation (multi-$Q$ order). The resulting patterns give rise to a topological Hall effect, a critical signal for identifying candidate materials.
		
Two routes to topological magnetic textures have been identified in bulk materials. Originally, transition metal compounds without inversion symmetry (non-centrosymmetric) were explored.  In these materials, Dzyaloshinskii-Moriya interaction can promote the swirling non-coplanar magnetic configurations\cite{Tokura2020_MagneticSkyrmionMaterials,Back2020_SkyrmionRoadmap}. An alternative route has been proposed in centrosymmetric materials. In these systems, non-coplanar magnetic textures arise from lattice frustration or competition between nearest and next nearest neighbor interactions\cite{Lin2016_GLTheorySkyrmionCentrosymmetricMagnets,Leonov2015_MultiplyPeriodicState+SkyrmionsFrustratedMagnets,Hayami2016_Bubble+SkyrmionCrystalsFrustratedEasyAxisMagnets,Okubo2012_MultiQ+SkyrmionTriagularLatticeAFM,Yambe2021_SkyrmionCentrosymmetricItinerantMagnetsWOMirror}. In metallic rare earth system like Gd$_{2}$PdSi$_{3}$, Gd$_{3}$Ru$_{4}$Al$_{12}$, and GdRu$_{2}$Si$_{2}$, the RKKY interaction is proposed to provide the competing couplings\cite{Wang2020_SkyrmionsFromRKKY,Hirschberger2019_SkyrmionLattice-Gd3Ru4Al12,Khanh2020_NanometricSkyrmionLatticeGdRu2Si2,Kurumaji2019_SkyrmionsGd2PdSi3,Spachmann2021_MagnetoelasticCouplingGd2PdSi3,Khanh2020_NanometricSkyrmionLatticeGdRu2Si2,Wang2021_TetCentrosymmMeron+Skyrmion+VortexCrystals,Mitsumoto2021_SkyrmionCrystalRKKYin2D}. These systems have attracted attention because the smaller size of the skyrmions textures could allow for significantly smaller devices.

\begin{figure}
	\includegraphics[width=8.6cm]{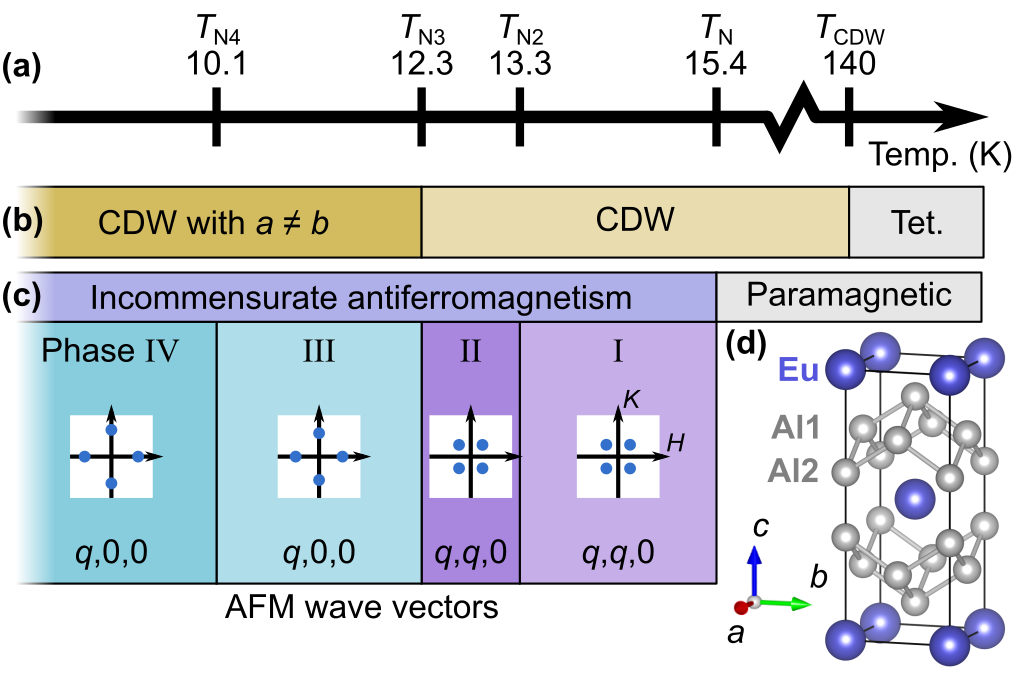}
	\caption{\label{fig:IntroFigure} 
		Summary of low temperature phases in EuAl$_{4}$. \textbf{(a)} Phase transition temperatures. \textbf{(b)} Progression of lattice phases. \textbf{(c)} The four zero-field magnetic phases and their roman numeral labels. The antiferromagnetic propagation vectors determined by Kaneko et al. \cite{Kaneko2021_EuAl4-SCNeutronDiffraction} for each phase are depicted as spots in the $HL0$ plane with values below. It is unknown if any of these phases have multi-$Q$ magnetic order. \textbf{(d)} Room-temperature structure of EuAl$_{4}$ rendered in VESTA\cite{Momma2011_Vesta3}. 
	}	
\end{figure}

Now we will introduce the skyrmion candidate EuAl$_{4}$ and its intricate low-temperature behavior. This tetragonal compound crystallizes with the BaAl$_{4}$/ThCr$_{2}$Si$_{2}$ structure type ($I4/mmm$ No.~139). This centrosymmetric structure (Fig.~\ref{fig:IntroFigure}\textbf{(d)}) is composed of square Eu nets separated by corrugated Al sheets. Members of the BaAl$_{4}$ family include $M$Al$_{4}$ and $M$Ga$_{4}$ with $M$ = Ca, Sr, Ba, and Eu. All are excellent metals\cite{Nakamura2014_EuGa4+EuAl4-CDW+QuantOsc,Nakamura2015_Transport+MagPropEuAl4+EuGa4,Nakamura2016_FermiSurf+CDWInSrAl4+EuAl4+BaAl4}. 

The zero-field behavior of EuAl$_{4}$ is summarized in Fig.~\ref{fig:IntroFigure} \cite{Araki2014_CDWinEuAl4,Nakamura2015_Transport+MagPropEuAl4+EuGa4}. Five transitions are observed in EuAl$_{4}$ below room temperature (panel \textbf{(a)}). Below $T_{\mathrm{CDW}}$, around 140\,K, a charge density wave (CDW) modulation appears\cite{Araki2014_CDWinEuAl4,Nakamura2015_Transport+MagPropEuAl4+EuGa4,Nakamura2016_FermiSurf+CDWInSrAl4+EuAl4+BaAl4,Ramakrishnan2022_OrthorhombicCDW-EuAl4}. SrAl$_{4}$ also hosts a CDW below 243\,K\cite{Nakamura2014_EuGa4+EuAl4-CDW+QuantOsc}. Shimomura et al.~made a detailed x-ray diffraction study of this lattice modulation in EuAl$_{4}$ and identified a modulation wave vector of $(0,0,0.18)$\,r.l.u.\cite{Shimomura2019_CDW-EuAl4}. More recent analysis of single crystal data by Ramakrishnan et al. suggest that an orthorhombic CDW modulation is the most likely\cite{Ramakrishnan2022_OrthorhombicCDW-EuAl4}. Despite broken tetragonal symmetry below $T_{\mathrm{CDW}}$ no peak splitting was observed by either diffraction study although, an in-plane shear mode should be allowed. Shimomura et al. observed distinct breaking of tetragonal symmetry below roughly 12\,K (around $T_{\mathrm{N3}}$). In this low-temperature range, (left box in Fig.~\ref{fig:IntroFigure}\textbf{(b)}) $a$ and $b$ differ by $>$0.1\%. 

Next, we introduce the magnetic phases in EuAl$_{4}$ (summarized in Fig.~\ref{fig:IntroFigure}\textbf{(c)}). Europium in this compound is divalent \cite{Nakamura2014_EuGa4+EuAl4-CDW+QuantOsc,Wickman1966_MossbauerHF+IS-MagOrderedEuCompounds} and therefore possesses a spin-only magnetic moment ($S=7/2$). Antiferromagnetic order develops at $T_{\mathrm{N}} = 15.4$\,K followed by three additional magnetic transitions on cooling\cite{Nakamura2014_EuGa4+EuAl4-CDW+QuantOsc,Nakamura2015_Transport+MagPropEuAl4+EuGa4,Shang2021_AHE-EuAl4}. Roman numerals are used throughout this paper to label the numerous magnetic phases. 

The neutron diffraction study by Kaneko et al.~revealed the four zero-field phases host incommensurate magnetic order\cite{Kaneko2021_EuAl4-SCNeutronDiffraction}. As depicted in Fig.~\ref{fig:IntroFigure}\textbf{(c)}, both phases I and II host incommensurate modulations with propagation vectors $(q,q,0)$ and $(q,-q,0)$. Below $T_{\mathrm{N3}} = 12.3$\,K a different incommensurate magnetic order is present. In phases III and IV the magnetic diffraction peaks were indexed to $(q,0,0)$ and $(0,q,0)$. The temperature dependence of CDW in Ref.~\cite{Shimomura2019_CDW-EuAl4} shows clear competition between the $(q,0,0)$ magnetic order and the CDW modulation. Finally, it is not clear if any of these phases have multi-$Q$ magnetic order.

The evolution of the four antiferromagnetic phases with magnetic field was examined in Ref.~\cite{Nakamura2015_Transport+MagPropEuAl4+EuGa4}. They report a curious series of metamagnetic transitions with increasing field along the [001] direction. Subsequently, Shang et al. reported that the last phase before the field polarized phase (phase 0) had an additional Hall contribution they suggested could be a topological contribution from a skyrmion phase\cite{Shang2021_AHE-EuAl4}. A temperature-field region with a similar Hall signal was observed in EuGa$_{4}$ and EuGa$_{2}$Al$_{2}$\cite{Zhang2021_GiantMagresist+TopoHE-EuGa4,Moya2021_EuGa2Al2-MagOrder+AHE+Neutron+XRD}. These phases with unusual Hall effects are reminiscent of the topological Hall phase observed in GdRu$_{2}$Si$_{2}$, which shares the same structure type and spin 7/2 moment\cite{Khanh2020_NanometricSkyrmionLatticeGdRu2Si2}. In this case, the Hall contribution was determined to arise from a skyrmion or meron\cite{Wang2021_TetCentrosymmMeron+Skyrmion+VortexCrystals} phase. It is natural to ask; does EuAl$_{4}$ host a skyrmion or meron phase in the region where the topological Hall effect is observed?

\begin{figure}
	\includegraphics[width=8.6cm]{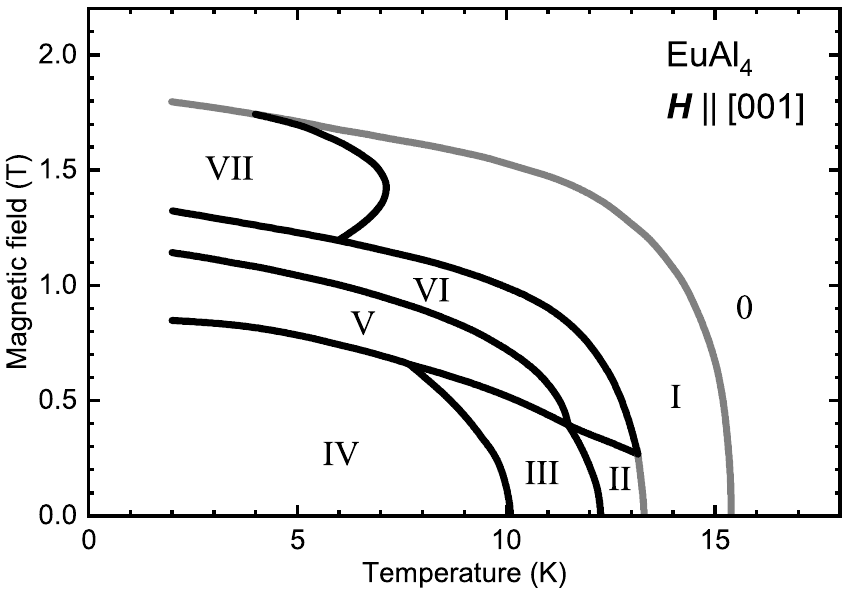}
	\caption{\label{fig:CartoonPhaseDiagram} Schematic phase diagram of magnetic phases in EuAl$_{4}$ for $c$-axis magnetic field determined in this study. Phase fields are labeled with roman numerals. Phase 0 is the paramagnetic or field polarized phase with the CDW modulation. Black and gray lines represent phase boundaries we believe are first and second order, respectively. The I-VII, II-VI, and III-V phase boundaries are new phase transitions proposed in this study.
	}
\end{figure}

In this paper we provide thermodynamic insights into the intricate magnetic phase diagram of EuAl$_{4}$ under $c$-axis magnetic field ($\bm{H}$\,$\parallel$\,$[001]$). The most important product of our investigation is a revised phase diagram for EuAl$_{4}$ (depicted in Fig.~\ref{fig:CartoonPhaseDiagram}). We clearly identify seven magnetic phases labeled with roman numerals. We also present evidence for which transitions are first or second order (black and gray lines in Fig.~\ref{fig:CartoonPhaseDiagram}, respectively). The subtle, first order I-VII transition is particularly significant. Phase VII corresponds to the $T,H$ region where the topological Hall effect is reported in Ref.~\cite{Shang2021_AHE-EuAl4}. The characteristics of this phase are suggestive of either a skyrmion or meron crystal. Our detailed phase diagram and insights into the individual phases will guide future investigations to uncover details of the interactions that lead to the rich magnetic behavior of EuAl$_{4}$.

The paper is organized as follows. First, in section \ref{sec:Methods} we describe how we obtained and characterized the EuAl$_{4}$ single crystals. In section \ref{sec:Results} we present our dilatometry, heat capacity, DC magnetization, AC magnetic susceptibility, and resonant ultrasound spectroscopy results. We examine the characteristic signatures of the CDW transition and zero-field magnetic phases in sections \ref{sec:Results_ChargeDensityWave} and \ref{sec:Results_ZeroFieldTrans}. Importantly, we identify three phase transitions not previously observed (secs.~\ref{sec:Results_LowFieldTrans} and \ref{sec:Results_HighFieldTrans}). 

In total, we identify seven magnetic phases in the $\bm{H}$\,$\parallel$\,$[001]$ phase diagram (Fig.~\ref{fig:CartoonPhaseDiagram}). In section \ref{sec:Discussion} we discuss the characteristics of these phases in greater detail. Multi-$Q$ skyrmion and meron crystals are closely tied to the tetragonal symmetry \cite{Wang2021_TetCentrosymmMeron+Skyrmion+VortexCrystals}. In section \ref{sec:Discussion_TetragonalPhases} we discuss which phases show strong distortions from tetragonal symmetry. In section \ref{sec:Discussion_Phases+Transition} we explore the character of the magnetic phases and transitions in light of our thermodynamic measurements. Next, we discuss what we know about the intriguing phase VII (sec.~\ref{sec:Discussion_NaturePhaseVII}). Finally, we discuss future directions of investigation in EuAl$_{4}$ (sec.~\ref{sec:Discussion_EuAl4Outlook}).

\section{Methods}
\label{sec:Methods}

\subsection{Growth}
\label{sec:Methods_growth}

EuAl$_{4}$ crystals were grown from a high-temperature aluminum-rich melt, as in previous works \cite{Nakamura2015_Transport+MagPropEuAl4+EuGa4,Tobash2006_SynthStructPropertiesSiDopedEuAl4+TmAlSi+LuAlSi,Stavinoha2018_CDWinEuGa4-EuAl4,StavinohaThesis2019,Nakamura2016_FermiSurf+CDWInSrAl4+EuAl4+BaAl4}. We roughly followed the method from Nakamura et al. starting with a Eu:Al = 1:9 atomic ratio\cite{Nakamura2015_Transport+MagPropEuAl4+EuGa4}. Eu pieces (Ames Laboratory, Materials Preparation Center 99.99+\%) and Al shot (Alfa Aesar 99.999\%) totaling 2.5\,g were loaded into one side of a 2\,mL alumina Canfield Crucible Set \cite{Canfield2016_CanfieldCrucibleSet}. The crucible set was sealed under 1/3\,atm argon in a fused silica ampoule.

The ampoule assembly was placed in a box furnace and heated to 900\textdegree C over 6\,h (150\,\textdegree C/h) and held for 12\,h to melt and homogenize the metals. Crystals were precipitated from the melt during a slow cool to 700\textdegree C over 100\,h (-2\,\textdegree C/h). To liberate the crystals from the remaining liquid the hot ampoule was removed from the furnace, inverted into a centrifuge, and spun.

\subsection{Products}
\label{sec:Methods_Products}

\begin{figure}
	\includegraphics[width=8.6cm]{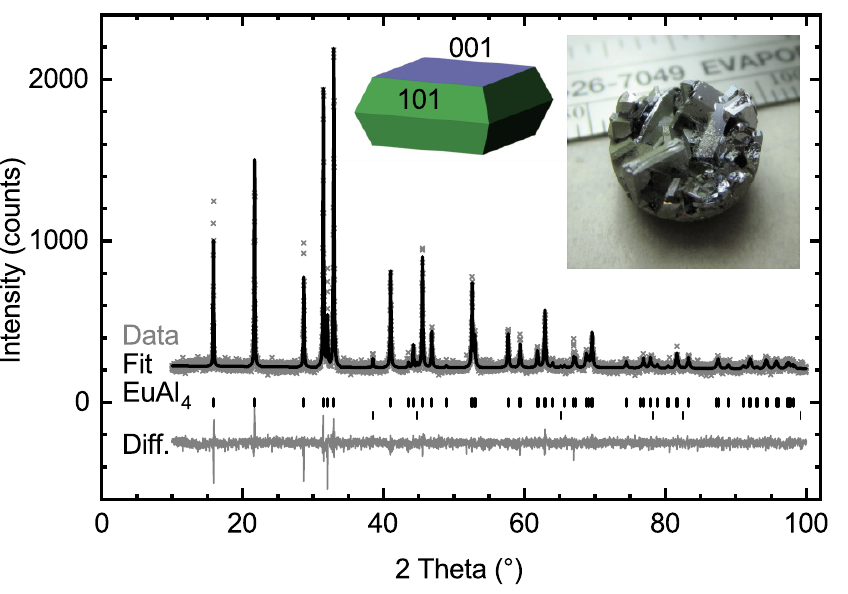}
	\caption{\label{fig:XRD} Powder x-ray diffraction pattern from ground EuAl$_{4}$ crystals. Rietveld refinement estimates majority EuAl$_{4}$ and about 3\,wt\% Al-metal (upper and lower rows of ticks, respectively). The middle inset depicts the observed to-go box crystal form rendered in JCrystal (http://www.jcrystal.com/jcrystal.html) with labeled faces. The right inset is a photo of an attractive cluster of metallic EuAl$_{4}$ exemplify their well-faceted, tabular habit. A millimeter scale sits behind.
	}
\end{figure}

The right-hand inset in Fig.~\ref{fig:XRD} shows a cluster of nicer EuAl$_{4}$ crystals obtain from this procedure. They look very similar to those of other BaAl$_{4}$-type aluminides\cite{Nakamura2014_EuGa4+EuAl4-CDW+QuantOsc,Nakamura2016_FermiSurf+CDWInSrAl4+EuAl4+BaAl4}. These silver-metallic crystals ranged from rounded, anhedral sub-millimeter grains up to blocky or tabular crystals 6\,mm wide and 2\,mm thick. These faceted grains adopted a clam-shell to-go box shape depicted in middle inset of Fig.~\ref{fig:XRD}. This shape is also called a square bifrustum and has rectangular basal [001] faces and trapezoidal $\left\lbrace 101 \right\rbrace$ faces (determined by x-ray diffraction). 

Clean crystals had mirror-like faces but many grains were coated in residual soft aluminum metal covered with a earthy crusty yellow-green phase we suspect to be Eu oxides or hydroxides. Most aluminum metal could be removed by an overnight soak in 0.6 M HCl or a 1.2\,M HNO$_{3}$ aqueous solution but, the crusty material remained. EuAl$_{4}$ was not visibly attacked by these acid solutions. The secondary phases were removed when crystals were polished to shape for our experimental investigations. These brittle crystals exhibit conchoidal fracture and are stable in air for many months.

Room temperature powder X-ray diffraction on ground crystals was preformed using PANalytical X'pert Pro diffractometer equipped with a Cu K$\alpha$ tube and an incident beam monochromator. FullProf was used for Rietveld refinement of the pattern (Fig.~\ref{fig:XRD}). The majority phase was EuAl$_{4}$ with the BaAl$_{4}$/ThCr$_{2}$Si$_{2}$ structure type (Fig.~\ref{fig:IntroFigure}\textbf{(d)}) with 3\,wt\% aluminum metal from the crystal surfaces. The refined structural parameters are consistent with those in previous works\cite{Nakamura2015_Transport+MagPropEuAl4+EuGa4,StavinohaThesis2019,Stavinoha2018_CDWinEuGa4-EuAl4,Zhang2013_GeDopedBaAl4Type}: $a = 4.39860(12)$\,\AA, $c = 11.1740(4)$\,\AA, and Al2 $z = 0.3874(5)$.

\subsection{Measurements}
\label{sec:Methods_Measurements}

One EuAl$_{4}$ crystal was carefully shaped into a rectangular prism for dilatometry and subsequently used for magnetometry and heat capacity measurements. The dimensions were $2.01 \times 1.06 \times 0.72$\,mm$^3$ along the $[100]$, $[010]$, and $[001]$ directions, respectively. 

Dilatometry measurements were obtain using the Quantum Design dilatometer option\cite{Martien2019_QD-Dilatometer} in a Quantum Design PPMS DynaCool system. The sample was mounted with magnetic field along the $\left[001\right]$ direction monitoring the length change along the longest direction, [100]. Dilation vs temperature was measured on cooling from 360 to 2\,K at 12\,mK/s. Detailed low-temperature thermal expansion data were obtained on warming and cooling between 2 and 20 K at 5\,mK/s under constant applied fields up to 4\,T. Dilation vs field data were obtained at constant temperatures from 2 to 18\,K by ramping field at 1\,mT/s between 0 and 3\,T.

Specific heat capacity ($C_{p}$) data were obtained with the heat capacity option on the same PPMS DynaCool system. The dilatometry sample was mounted on the platform with Apiezon N grease with magnetic field along the [001] direction. Heat capacity was measured using both a standard 2\% temperature rise and the ``large-pulse'' method with a 20\% rise. Post processing the raw data from the latter method in Quantum design Multiview software allows us to obtain a densely spaced $C_{p}(T)$ data as the sample warms and cools using the single-slope option. This allows us to capture the evolution of the closely spaced first order transitions in EuAl$_{4}$ with temperature and field. Most of the data presented in this paper was obtained with the long-pulse method but $C_{p}$ values are very consistent with the standard method at temperatures away from phase transitions (see Fig.~\ref{fig:ZeroFieldPhases}c).

Magnetometry measurements were carried out in a Quantum Design MPMS3 system with field along the the $c$-axis. The dilatometry crystal was mounted to a fused silica rod with GE-varnish with field along $\left[001\right]$. DC magnetization vs temperature data was measured in constant field up to 2.2\,T between 2 and 25\,K with a rate of 5\,mK/s. This means that temperature sweeps in magnetometry and dilatometry are directly comparable. DC magnetization vs field data was measured at constant temperatures between 2 and 18\,K by stabilizing at each field up to 2.5\,T. AC magnetometry was also carried out in the MPMS3 using a 257.67\,Hz, 0.5\,mT drive field. Temperature dependent data was obtained on cooling from 20 to 2\,K at 1.7\,mK/s at constant fields up to 2\,T.

Resonant ultrasound spectroscopy (RUS) measurements were performed as a function of temperature and applied magnetic field using a custom-built RUS probe compatible with the Quantum Design PPMS. The Alamo Creek Engineering (ACE) RUS008 system\cite{Balakirev2019_RUS-EssentialToolbox} was used for signal generation and detection, driven by a homemade Python implementation of the ACE software. The EuAl$_{4}$ sample was approximately rectangular parallelepiped shaped with nominal dimensions $1.292\times 1.902\times 0.745$\,mm$^3$ (short length along the $\left[001\right]$ direction). A general measurement scheme was employed for all temperature- and applied-field-dependent measurements: frequency scans were recorded continuously over the 0.7-1.2\,MHz range with a step size of 20\,Hz and dwell time of 4\,ms. Temperature and field were ramped continuously at 3.3\,mK/s and 0.2\,mT/s, respectively; thus, each RUS scan spanned ranges of about 0.3\,K and 20\,mT, respectively.

$^{151}$Eu M\"ossbauer spectra were collected between 20 and 170\,K on a 95\,mg/cm$^{2}$ powder sample of EuAl$_{4}$ using a 50\,mCi $^{151}$SmF$_{3}$ source at ambient temperature. The sample was placed in a Janis SHI-850 closed cycle cryostat. A Wissel GmbH drive in constant acceleration mode was used in the $\pm$30\,mm/s range and calibrated by measuring a room temperature spectrum of $\alpha$-iron. The isomer shift is reported relative to the $^{151}$SmF$_{3}$ source. A Tl@NaI scintillator (Ametek) was used as detector.

$^{151}$Eu Nuclear resonant inelastic x-ray scattering (NRIXS) spectra using 21.54\,keV synchrotron radiation\cite{Leupold1996_NuclearResonanceScattering151Eu} were measured at the 3-ID beamline of the Advanced Photon Source, Argonne National Laboratory with 0.8\,meV resolution. NRIXS measures the phonon assisted nuclear resonant absorption of radiation which yields the partial vibrational inelastic scattering function, $S(E)$, for the resonant element. The scattering function is directionally projected on the direction of the incident beam in case of measurements on a single crystal\cite{Kohn1998_NuclearResonantInelasticAbsorptionOfSynchrotronRadiationSingleCrystal}. Here, we have utilized two, roughly 100\,\textmu m sized, single crystals of EuAl$_{4}$, respectively oriented with the beam parallel and perpendicular the $x$-axis. The crystals were glued on a diamond and placed in a miniature cryostat described in Ref.~\cite{Zhao2017_DiamondAnvilCell+CryostatForNuclearResonantScattering}.

\section{Results}
\label{sec:Results}

In this section we will present thermodynamic data revealing the numerous phase transitions in EuAl$_{4}$. We not only observe the previously identified charge density wave (section~\ref{sec:Results_ChargeDensityWave}) and four zero-field magnetic transitions (section~\ref{sec:Results_ZeroFieldTrans}) but also three transitions not observed previously (sections~\ref{sec:Results_LowFieldTrans} and \ref{sec:Results_HighFieldTrans}). We note that only select magnetization data are presented here, but a very large number of fields and temperatures were examined to produce the points utilized to draw smooth lines in the presented phase diagrams.  Phase diagrams containing these points are shown at the end of this section.

\subsection{Charge density wave}
\label{sec:Results_ChargeDensityWave}

\begin{figure}
	\includegraphics[width=8.6cm]{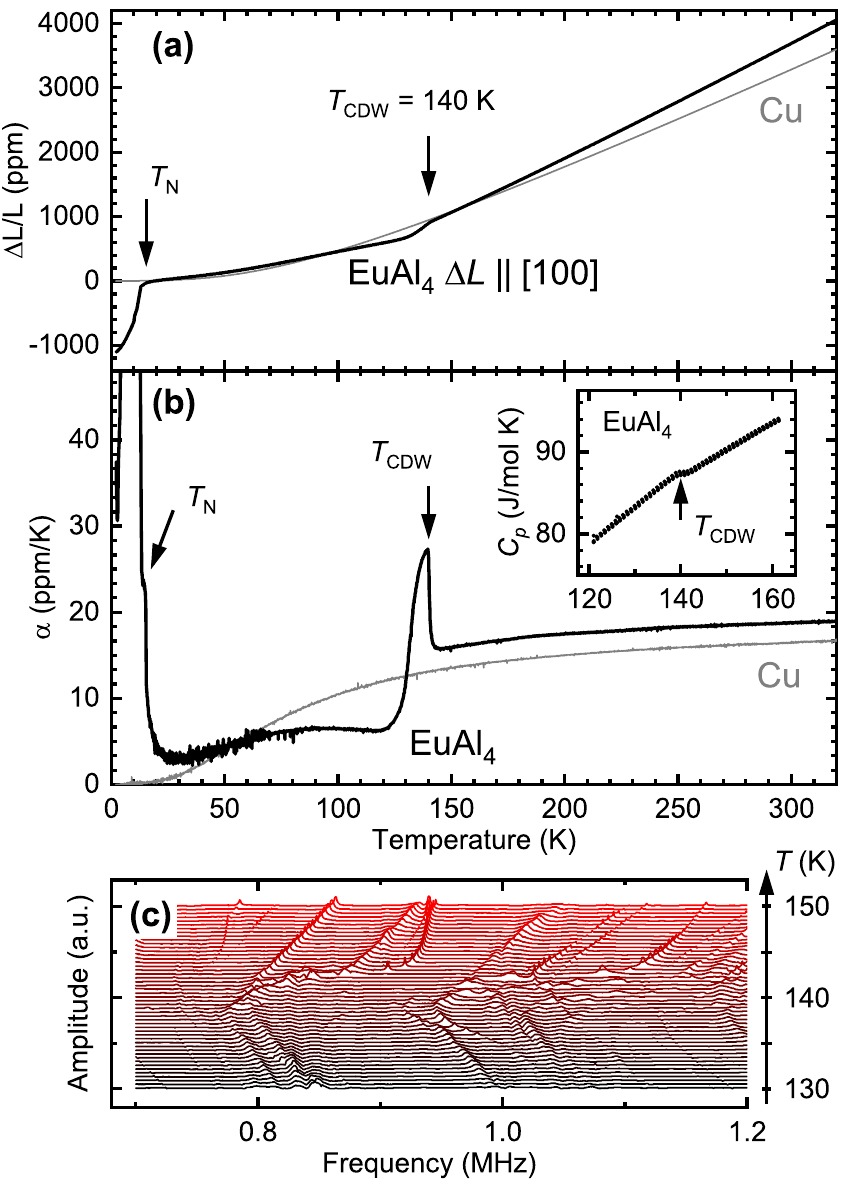}
	\caption{\label{fig:CDW} Thermodynamic signatures of the charge density wave (CDW) transition in EuAl$_{4}$. \textbf{(a)} The thermal expansion, $\Delta L/L$, of EuAl$_{4}$ and copper metal measured on cooling. \textbf{(b)} The coefficient of thermal expansion ($\alpha$) for EuAl$_{4}$ and copper metal. The inset shows the specific heat capacity jump at $T_\textrm{CDW}$ consistent with a second order phase transition. \textbf{(c)} Resonant ultrasound spectroscopy spectra at temperatures spanning the CDW transition measured on warming. Peaks reflect mechanical resonances of the sample.
	}
\end{figure}

Previous investigations of EuAl$_{4}$ have observed a charge density wave (CDW) below about 140\,K \cite{Shimomura2019_CDW-EuAl4,Kaneko2021_EuAl4-SCNeutronDiffraction,Nakamura2015_Transport+MagPropEuAl4+EuGa4,Moya2021_EuGa2Al2-MagOrder+AHE+Neutron+XRD,Ramakrishnan2022_OrthorhombicCDW-EuAl4}. Figure~\ref{fig:CDW} presents thermodynamic signatures of this transition. In panel \textbf{(a)} we present the thermal expansion EuAl$_{4}$ along the $\left[100\right]$ direction compared to polycrystalline copper. Plots of the thermal expansion coefficient vs temperature of the metals, $\alpha(T) = \frac{1}{L_0} \frac{dL}{dT}$, are presented in panel \textbf{(b)}. Although the length change of EuAl$_{4}$ is comparable to Cu down to 150\,K, it shows dramatic features below this temperature. The CDW transition is marked by a kink in panel \textbf{(a)} and a step-like increase in $\alpha_a$ by 73\% on cooling through $T_\textrm{CDW} = 140$\,K (indicated by the arrow). A step in $\alpha_{a}(T)$ is characteristic of a second order phase transition. Cooling below this transition, $\alpha_a$ smoothly falls to a value 60\% smaller than just above $T_\textrm{CDW}$. EuAl$_{4}$ shows dramatic changes in length below the magnetic ordering temperature, $T_{\mathrm{N}}$. Below this temperature $\alpha_{a}(T)$ rises steeply. We will examine this low temperature behavior in more detail in section~\ref{sec:Results_ZeroFieldTrans}.

The inset of Fig.~\ref{fig:CDW}\textbf{(c)} presents the specific heat capacity of EuAl$_{4}$ across the CDW transition. The step-like feature corroborates the second order character of the CDW transition identified in thermal expansion. 

Resonant ultrasound spectroscopy (RUS) measurements also observe a clear signature of the CDW transition. Figure~\ref{fig:CDW}\textbf{(c)} presents a series of RUS spectra at a range of temperatures spanning $T_\textrm{CDW}$. Mechanical resonance modes of the EuAl$_{4}$ sample give a larger vibration amplitude near the resonant frequency, observed as peaks in each spectrum. The resonant frequencies shift as the elastic moduli evolve with temperature. Critically, when moduli stiffen, the resonant frequencies increase and peaks shift to the right.

In Fig.~\ref{fig:CDW}\textbf{(c)}, note that peaks in the spectra move to lower frequencies as we approach $T_\textrm{CDW} = 140$\,K from above. Below $T_\textrm{CDW}$ we observe the opposite trend; peak frequencies increase on cooling. This indicates that the elastic modes of the crystal are softening on cooling toward CDW transition and stiffening again below. This evolution of elasticity reflects the coupling of the CDW to the bulk elastic modes. In addition, most of the resonant mode streaks are continuous through the phase transition. This also suggests that $T_\textrm{CDW}$ is a second order transition because we observe an abrupt change in slope of the elastic moduli vs temperature\cite{Rehwald1973_StudyPhaseTransitionsUltrasound}. 

We observed the CDW superlattice peaks using single crystal x-ray diffraction at 100\,K but we don't present it here. We indexed these with a wavevector of (0 0 0.1822(15))\,r.l.u., in good agreement with previous x-ray\cite{Shimomura2019_CDW-EuAl4,Moya2021_EuGa2Al2-MagOrder+AHE+Neutron+XRD} and neutron\cite{Kaneko2021_EuAl4-SCNeutronDiffraction} studies. 

We do not observe a signature of the CDW transition in the $^{151}$Eu M\"ossbauer spectral parameters within experimental error bars (Appendix \ref{sec:Appendix_Mossbauer}). This suggest that the CDW modulation has a more significant impact the Al sub-lattice rather than the Eu sub-lattice. NRIXS does offer a clue about the Eu atoms in the CDW phase. Below $T_\mathrm{CDW}$, the Eu atoms show a 30\% larger atomic displacement parameters along $c$ than in the $ab$-plane (Appendix \ref{sec:Appendix_Mossbauer}). Near room temperature, the dynamic displacement parameters are more isotropic. This likely reflects a change from dynamic displacements of Eu perpendicular to [001] above $T_\mathrm{CDW}$, to static displacements in the CDW phase.

Our thermodynamic data shows clear indications of the CDW transition observed in previous studies. In addition, we provide evidence that the transition is second order. Finally, we observe that the CDW order has a significant impact on the lattice thermal expansion and elasticity. 

\subsection{Magnetic transitions in zero field}
\label{sec:Results_ZeroFieldTrans}

\begin{figure}
	\includegraphics{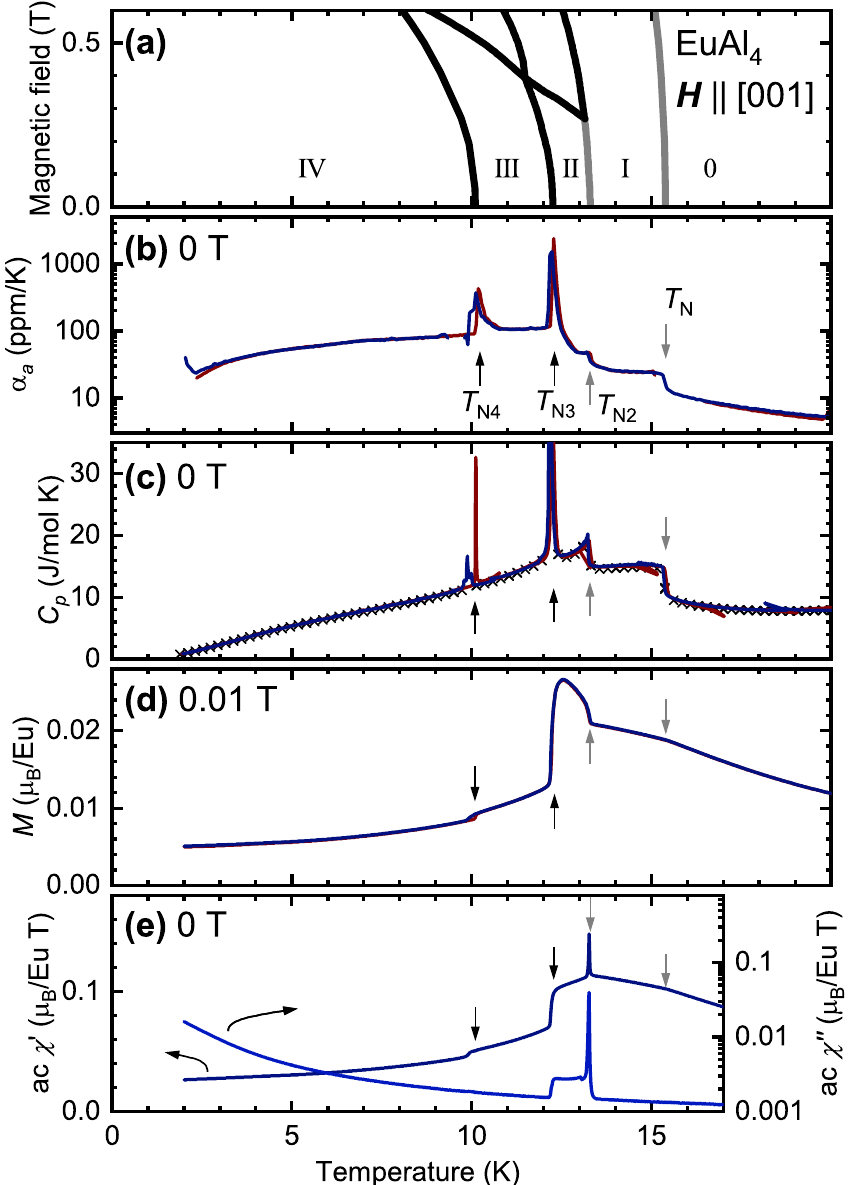}
	\caption{\label{fig:ZeroFieldPhases} Zero field magnetic phases of EuAl$_{4}$. \textbf{(a)} Low-field section of the phase diagram showing labeled phase fields. \textbf{(b)} Thermal expansion coefficient, $\alpha_a(T)$, measured on heating (red) and cooling (blue) showing all four transitions. \textbf{(c)} Zero field heat capacity data. Red and blue curves are continuous heat capacity data derived from the heating and cooling parts of the long-pulse heat capacity measurement (see section \ref{sec:Methods_Measurements}). $\times$'s are heat capacity values from the standard heat capacity measurement demonstrating nice agreement with the long-pulse data. \textbf{(d)} DC-magnetization measurement along [001] at 0.01\,T showing all four transitions on heating (red) and cooling (blue). \textbf{(e)} AC magnetic susceptibility data taken on cooling at zero field. The in-phase ($\chi'$) and out of phase ($\chi''$) parts are presented in dark and light blue.
	}
\end{figure}

Here, we will examine the succession of magnetic orders at low temperature in zero magnetic field. As mentioned in the introduction, EuAl$_{4}$ undergoes four magnetic transitions on cooling in zero field (Fig.~\ref{fig:IntroFigure}). We observe all four in our thermodynamic measurements shown in Fig.~\ref{fig:ZeroFieldPhases}. Panel \textbf{(a)} depicts the phase transitions of EuAl$_{4}$ under small magnetic fields along [001] for context. 

Figure \ref{fig:ZeroFieldPhases}\textbf{(b)} presents the low temperature $\left[100\right]$ coefficient of thermal expansion ($\alpha_a$) on warming and cooling (red and blue, respectively). First, note the humongous values of $\alpha_a$ within the phases below $T_\textrm{N}$; ranging from 20 to 100\,ppm/K. This is dramatically larger than Cu or Al with $\alpha$'s of 0.001 to 0.3\,ppm/K within this temperature range\cite{Kroeger1977_CTE-Cu+Al}. $T_\textrm{N}$ and $T_\textrm{N2}$ are evident as steps in $\alpha_a(T)$ marked by gray arrows. These are suggestive of second order transitions. In addition, large peaks in $\alpha_a(T)$ appear at $T_\textrm{N3} = 12.3$\,K and $T_\textrm{N4} = 10.1$\,K indicative of first order phase transitions. These are accompanied by relative length changes of 230 and 80\,ppm, respectively. Shimomura et al.~observed tetragonal symmetry breaking at $T_\textrm{N3}$ which accounts the large strain we observe\cite{Shimomura2019_CDW-EuAl4}. We will discuss the symmetry implication of the these large strains in section~\ref{sec:Discussion_TetragonalPhases}. Clearly, magnetic order is strongly coupled to lattice strain.

The low temperature heat capacity of EuAl$_{4}$ is presented in Fig.~\ref{fig:ZeroFieldPhases}\textbf{(c)}. Our $C_{p}(T)$ data are in excellent quantitative agreement with those presented in Ref.~\cite{Nakamura2015_Transport+MagPropEuAl4+EuGa4}. Red and blue curves depict $C_{p}(T)$ curves from warming and cooling parts of a 20\% temperature pulse obtained by the large-pulse method. These values are nearly identical to the values of $C_{p}$ obtained by the standard relaxation time approach with a 2\% temperature rise (plotted as $\times$'s). The large-pulse slope analysis allows a high density of points and, critically, treatment of first-order transitions. The features in the heat capacity data closely track the features of the thermal expansion plot with steps at $T_\textrm{N}$ and $T_\textrm{N2}$ and peaks at $T_\textrm{N3}$ and $T_\textrm{N4}$ once again marked by arrows.

Next, we will consider the DC-magnetization data in Fig.~\ref{fig:ZeroFieldPhases}\textbf{(d)}. This was measured at a small field (0.01\,T) to obtain an estimate of the zero-field magnetic susceptibility. Magnetization rises on cooling to $T_\textrm{N}$ which is marked by a slope change. There is an abrupt increase at $T_\textrm{N2}$ with an elevated value of $M$ in phase~II down to $T_\textrm{N3}$ (also observed at the same field in Ref.~\cite{Shang2021_AHE-EuAl4}). Both $T_\textrm{N3}$ and $T_\textrm{N4}$ are characterized by drops in magnetization on cooling. Data taken on warming and cooling (red and blue, respectively) are nearly identical with only a weak hysteresis visible at $T_\textrm{N4}$. 

AC-magnetic susceptibility (measured on cooling) tells a subtly different story (Fig.~\ref{fig:ZeroFieldPhases}\textbf{(e)}). The in-phase susceptibility, $\chi'(T)$, has slope changes at $T_\textrm{N}$ and $T_\textrm{N2}$ and steps $T_\textrm{N3}$ and $T_\textrm{N4}$. Curiously, there is a discrepancy between the AC and DC susceptibility in phase~II. $M(T)$ shows a rounded, negatively sloped plateau in this regime whereas $\chi'(T)$ shows a sharp peak at $T_\textrm{N2}$ and a positive sloped region just below. The imaginary part of the AC-susceptibility ($\chi''(T)$) also shows different behavior in phase II. There a sharp peak at $T_\textrm{N2}$ and a flat topped plateau down to $T_\textrm{N3}$. We will examine the discrepancy between the DC and AC magnetic data in phase II in section \ref{sec:Discussion_I-II-transition}. The $\chi'(T)$ features at $T_\textrm{N3}$ and $T_\textrm{N4}$ mirror the features observed in $M(T)$ suggestive of first order transitions.

We clearly detect the four previously reported magnetic transitions in EuAl$_{4}$ at zero field. The N\'eel transition ($T_\textrm{N}$) has second order character in all the measurements. The nature of the transition at $T_\textrm{N2}$ is less clear (see sec.~\ref{sec:Discussion_I-II-transition}). $T_\textrm{N3}$ and $T_\textrm{N4}$ are both clearly first order transitions indicated by the peaks in $\alpha_c(T)$ and $C_p(T)$ and steps in $M(T)$ and have a thermal hysteresis of about 0.1\,K. These designations are consistent with previous data on EuAl$_{4}$ \cite{Nakamura2015_Transport+MagPropEuAl4+EuGa4,Shang2021_AHE-EuAl4}.

\subsection{Magnetic transitions at low field}
\label{sec:Results_LowFieldTrans}

\begin{figure}
	\includegraphics{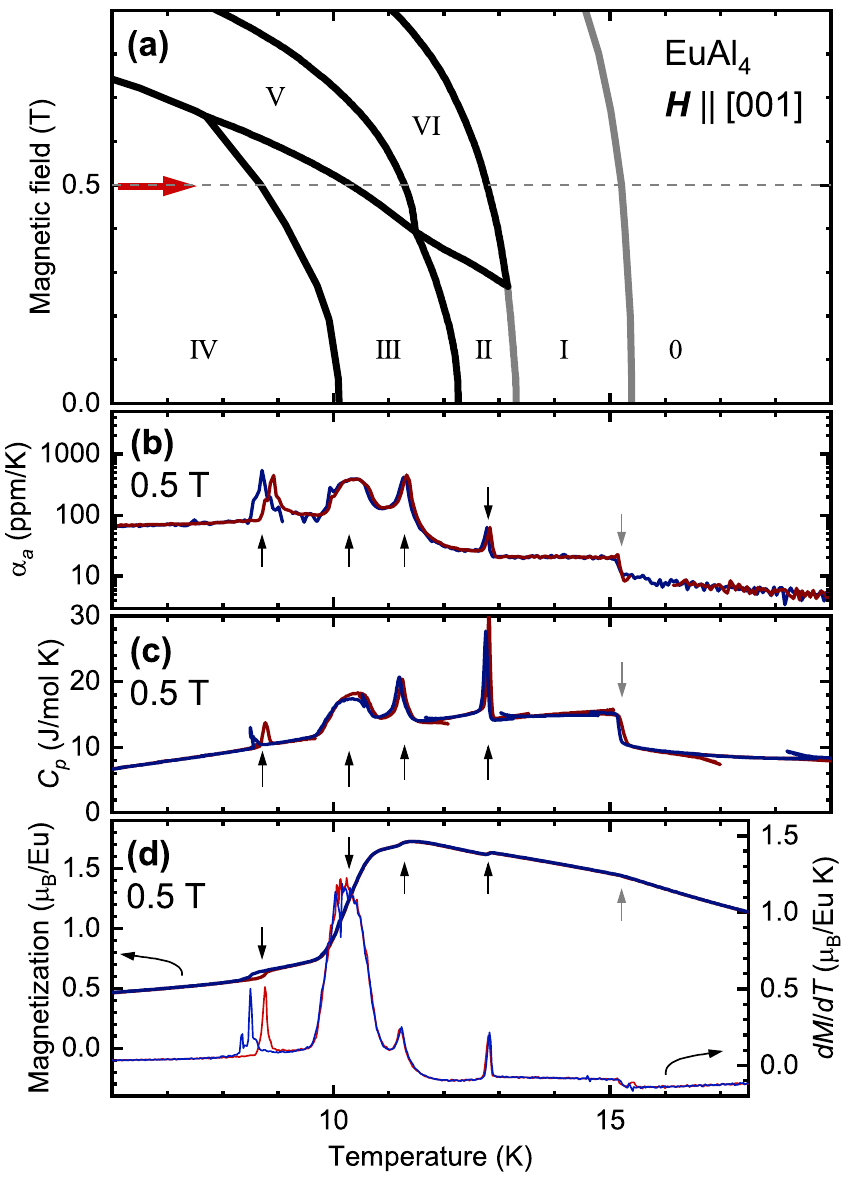}
	\caption{\label{fig:LowFieldPhasesVsTemp} New low field transition evident 0.5\,T measurements. \textbf{(a)} Cropped phase diagram showing the phase fields in the region of interest. The arrow and horizontal dashed line place the measurements in context. \textbf{(b)} [100] thermal expansion data of EuAl$_{4}$ at 0.5\,T taken on warming (red) and cooling (blue). Five transitions are observed and marked with arrows. Note the log scale. \textbf{(c)} 0.5\,T heat capacity data obtained using the long pulse method. Red and blue curves represent $C_p$ values obtained from the warming and cooing parts of each pulse, respectively. \textbf{(d)} DC magnetization data (thicker curves) clearly revealing all five transitions on warming and cooling. The temperature derivatives, $\frac{dM}{dT}(T)$, are presented as thinner lines with the right hand axis.
	}
\end{figure}

\begin{figure*}
	\includegraphics{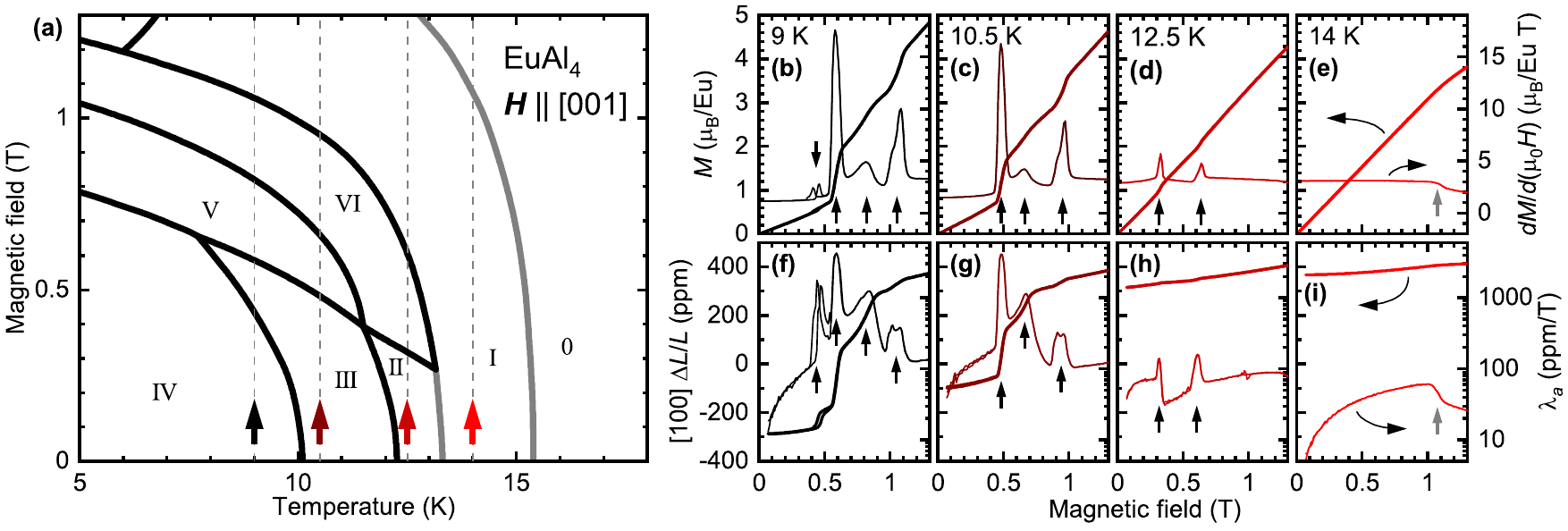}
	\caption{\label{fig:LowFieldPhasesVsH} Evidence for low-field phase transitions in field-dependent measurements. \textbf{(a)} Phase diagram showing the labeled phase fields. The arrows and vertical dashed lines reveal the isothermal cuts that correspond to the magnetic and dilation data on the right. \textbf{(b)}-\textbf{(e)} Thick and thin curves represent the field dependence of magnetization, $M$, and its derivative, $\frac{dM}{d(\mu_0 H)}$, at four temperatures. Black and gray arrow denote the first and second order transitions. \textbf{(f)}-\textbf{(i)} Thick and thin curves present the field dependent dilation ($\frac{\Delta L}{L}$) and magnetostriction coefficient ($\lambda_a$ = $\frac{1}{L_{0}} \frac{dL}{d(\mu_0 H)}$) as a function of field. Note that curves for both increasing and decreasing field are plotted in panels~\textbf{(b)}-\textbf{(i)}. Only the III-IV transition at 0.43\,T in the 9\,K plots show any significant hysteresis.
	}
\end{figure*}

We now turn our attention to the magnetic phases in EuAl$_{4}$ at low magnetic fields. Within this regime, we observe new transitions between 0.3 and 0.5\,T not previously reported.  Figures~\ref{fig:LowFieldPhasesVsTemp} and \ref{fig:LowFieldPhasesVsH} present our evidence for transitions between phases II and VI and phases III and V.

Figure~\ref{fig:LowFieldPhasesVsTemp} presents thermodynamic measurements at 0.5\,T showing evidence for the new transition between phases III and V centered at 10.3\,K. We observe 5 transitions in the thermal expansion, heat capacity and DC-magnetization measurements (panels \textbf{(b)}, \textbf{(c)}, and \textbf{(d)}) at this field (marked with arrows). The horizontal dashed line in panel~\textbf{(a)} places these transitions in context of the phase diagram. At this field, only the magnetic ordering transition, $T_\textrm{N}$, has second-order character as evidenced by steps in $\alpha_c(T)$, $C_p(T)$ and $\frac{dM}{dT}(T)$. 

For fields below 0.2\,T, the transition at $T_\textrm{N2}$ has predominant second order character (see Fig.~\ref{fig:ZeroFieldPhases}). At higher fields, the transition takes on first order character like that at 12.8\,K in 0.5\,T plots (Fig.~\ref{fig:LowFieldPhasesVsTemp}\textbf{(b)}-\textbf{(d)}). This change coincides with the triple point between phases I, II and VI. Further evidence for the two distinct phase fields for phases II to VI and the transition between them will be presented later in this section.

At lower temperatures, we see three additional first order transitions at 11.3, 10.3, and near 8.7\,K in our 0.5\,T data sets. These appear as peaks in $\alpha_c(T)$, $C_p(T)$ and $\frac{dM}{dT}(T)$. The new transition between phases III and V a 10.3\,K transitions stands out here as unusually broad in all three measurements (see sec.~\ref{sec:Discussion_II-VI_III-V-transitions}). Despite this, the transition has limited thermal hysteresis. This feature also appears in the field dependent data in Fig.~\ref{fig:LowFieldPhasesVsH} discussed below. The lowest temperature III-IV transition at 0.5\,T has a relatively strong thermal hysteresis of about 0.2\,K just like the corresponding transition at zero field, $T_\textrm{N4}$. 

Figure~\ref{fig:LowFieldPhasesVsH} provides further evidence of the new transitions. Panel~\textbf{(a)} depicts four isothermal sections through the phase diagram. Field dependent DC-magnetization and $a$-axis dilation data are presented on the right for each temperature (panels~\textbf{(b)}-\textbf{(e)} and \textbf{(f)}-\textbf{(i)}, respectively). Thinner lines represent the field derivatives of quantities, $\frac{dM}{d(\mu_0 H)}$, and the magnetostriction coefficient, $\lambda_a$ = $\frac{1}{L_{0}} \frac{dL}{d(\mu_0 H)}$. Note that the phase boundaries in the phase diagram in panel~\textbf{(a)} are reflected in the distinctive features in the field dependent data to the right marked with arrows.   

First we will consider how the III-IV phase boundary appears in these field dependent data. At 9\,K, the first transition appears as a hysteretic jump around 0.43\,T in the magnetization and dilation plots (as well as peaks in their derivatives). This feature is the continuation of the first order $T_\textrm{N4}$ transition observed in Figs.~\ref{fig:ZeroFieldPhases} and \ref{fig:LowFieldPhasesVsTemp}. As in those cases, the transition shows relatively strong hysteresis.

Next, we turn to the new III-V phase transition observed at 8.7\,K in Fig.~\ref{fig:LowFieldPhasesVsTemp}. It is clearly observed in the 9 and 10.5\,K data at 0.59 and 0.48\,T in Fig.~\ref{fig:LowFieldPhasesVsH}, respectively. On increasing field, the transition manifests as a strong step up in $M(\mu_0 H)$ and a significant extension of the crystal (1.2\,$\mu_\mathrm{B}/\mathrm{Eu}$ and 250\,ppm at 9\,K, respectively). These dramatic changes in thermodynamic properties in Figs.~\ref{fig:LowFieldPhasesVsTemp} and \ref{fig:LowFieldPhasesVsH} clearly suggest that phases III and V are distinct phases separated by a first order transition. Section~\ref{sec:Discussion_II-VI_III-V-transitions} provides a deeper discussion of this phase transition.

At higher fields, we can see the signature of the transition between phases V and VI. The transition appears in the 9 and 10.5\,K field dependent data as broad steeper region in the magnetization and dilation curves centered at 0.82 and 0.66\,T, respectively. We suspect that the V-VI transition is first order based on its character in temperature dependent measurements. In particular, the peaks observed in $\alpha_a (T)$, $C_p (T)$ and $\frac{dM}{dT}(T)$ in Fig.~\ref{fig:LowFieldPhasesVsTemp} clearly suggest a first order transition.

The I-VI phase boundary appears in phase diagrams already reported\cite{Shang2021_AHE-EuAl4,Nakamura2015_Transport+MagPropEuAl4+EuGa4}. In addition to the feature at 12.8\,K in Fig.~\ref{fig:LowFieldPhasesVsTemp}, we observe the I-VI transition in our field dependent results in Fig.~\ref{fig:LowFieldPhasesVsH}. Note the steps at 1.05, 0.95, and 0.64\,T in our $M(\mu_0 H)$ and $\Delta L/L$ plots measured at 9, 10.5 and 12.5\,K, respectively. Curiously, $\frac{dM}{d(\mu_0 H)}$ and $\lambda_a$ reveal double-peaks at this field at the lower two temperatures. This is particularly evident in the dilatometry results (panels~\textbf{(f)} and \textbf{(g)}) and AC-susceptibility measurement (not shown) and is discussed in greater detail in section \ref{sec:Discussion_I-VI-transition}. 

In addition to the new III-V phase boundary we propose above, our results also reveal a transition around 0.35\,T between phases II and VI. The field dependent data at 12.5\,K in Fig.~\ref{fig:LowFieldPhasesVsH}\textbf{(d)} shows a subtle jump of magnetization and sample length at 0.32\,T. This produces a sharp peak in $\frac{dM}{d(\mu_0 H)}$ and $\lambda_a$ plots with no obvious hysteresis. In addition to this feature, we also observe a change in the magnetostriction coefficient across the boundary. $\lambda_a$ falls from 70\,ppm/T in phase II to 34\,ppm/T in phase VI. As we noted above, there is also a change from second order I-II phase boundary to first order in the I-VI boundary within this field range. This all points to phases II and VI being distinct phases separated by previously unreported transition near 0.35\,T. See section \ref{sec:Discussion_II-VI_III-V-transitions} for more details.

Finally, we observe the transition between phase I and the paramagnetic/field-polarized phase. This is evident as a change in slope $M(\mu_0 H)$ and $\Delta L/L$ indicating second order character at all temperatures. Within the field range presented, we only see this feature in the 14\,K data (Fig.~\ref{fig:LowFieldPhasesVsH}\textbf{(e)} and \textbf{(i)}) and is most easily seen as a step in the field derivatives at 1.07\,T. 

Our thermodynamic data clearly reveal two new phase transitions in EuAl$_{4}$ in addition to those previously reported for magnetic fields along $[001]$ less than 1\,T. The results presented in Figs.~\ref{fig:LowFieldPhasesVsTemp} and \ref{fig:LowFieldPhasesVsH} provide strong evidence for a boundary between the phases III and V as well as phases II and VI. In section~\ref{sec:Discussion_II-VI_III-V-transitions}, we will discuss the clues our detailed measurements also offer about the nature all the phases.

\subsection{Magnetic transitions at high field}
\label{sec:Results_HighFieldTrans}

\begin{figure}
	\includegraphics{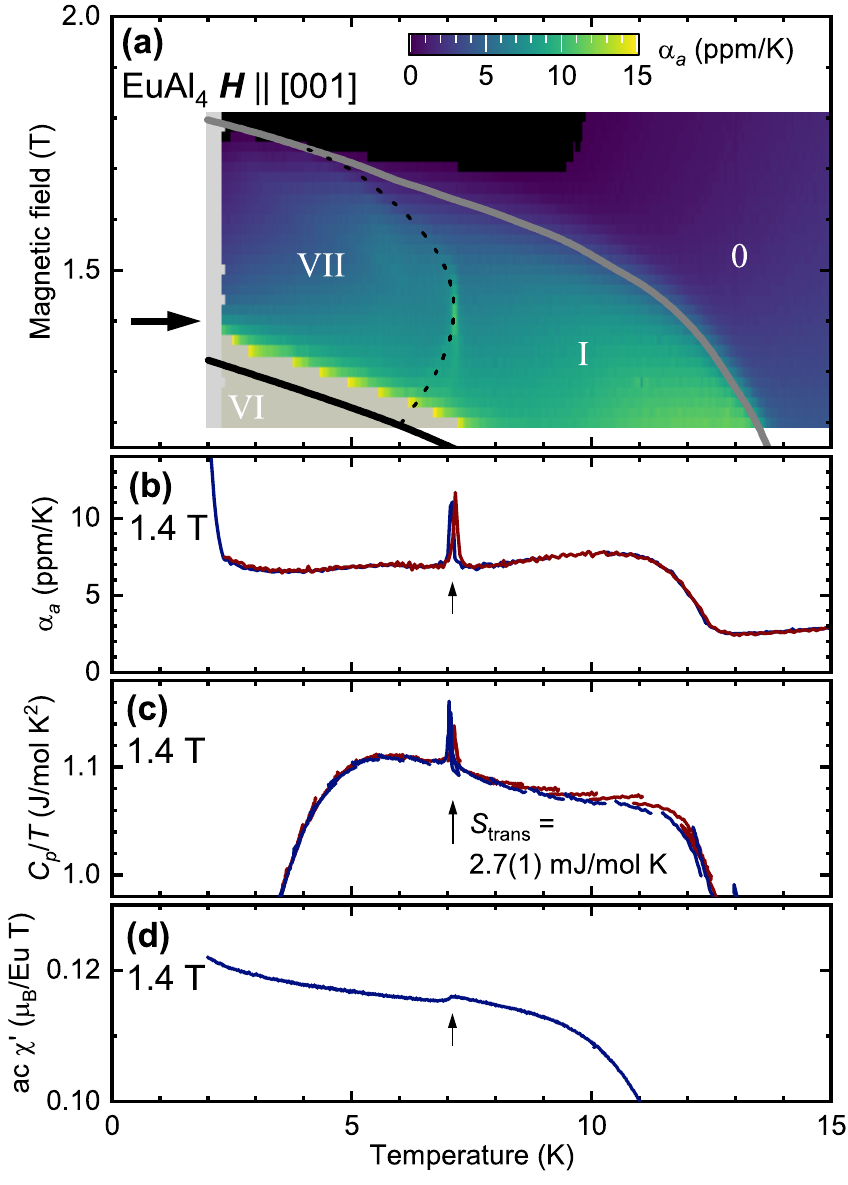}
	\caption{\label{fig:HighFieldPhases} Evidence for phase VII. \textbf{(a)} A color map representing the thermal expansion coefficient, $\alpha_a(T,H)$, with the phase boundaries overlain. Data was taking on cooling and brighter colors represent larger CTE (black and gray are out of range). The dashed line traces peaks in $\alpha_a(T,H)$ we interpret as a transition between phases I and VII. The arrow on the left emphasizes the 1.4\,T cut presented in the lower panels. \textbf{(b)} The thermal expansion data taken on heating (red) and cooling (blue) at 1.4\,T. \textbf{(c)} Heat capacity data taking using the long-pulse method. Red and blue curves present data derived from the warming and cooling segments of each heat-pulse measurement. \textbf{(d)} AC magnetic susceptibility with a static DC field of 1.4\,T taken on cooling displays a step-like feature at the I-VII transition.
	}
\end{figure}

In addition to the transitions proposed in the previous section, we identify an additional phase in EuAl$_{4}$ just below the saturation field. We observe a first order transition around 7\,K for $c$-axis fields between 1.3 and 1.8\,T. This phase VII region is important because this is the regime where a topological Hall contribution is proposed by Shang et al.\cite{Shang2021_AHE-EuAl4}. In this section we will introduce our evidence for a distinct new phase, phase VII.

Figure~\ref{fig:HighFieldPhases} depicts the extent of phase VII in EuAl$_{4}$ and evidence for its transition to phase I. Panel~\textbf{(a)} depicts a color map representing the [100] thermal expansion coefficient of EuAl$_{4}$ taken on cooling at constant field. Lighter colors represent higher values of $\alpha_a$. In this map we clearly observe the transition from the paramagnetic/field-polarized phase traced by the gray line. This second order transition manifests as a step in $\alpha_a(T,H)$ which appears in the color map as relatively abrupt change in color.

The most important feature in this plot is the light colored, curving line highlighted by a black dotted line. This corresponds to a series of peaks in $\alpha_a(T)$ that evolve smoothly with field between 5 and 7\,K. This feature separates phase fields I and VII. It is bounded at lower field by phase VI and merges with the saturation field near 1.75\,T.

To investigate this transition we will examine measurements taken at 1.4\,T (noted by the arrow in panel~\textbf{(a)}). Figure~\ref{fig:HighFieldPhases}\textbf{(b)} presents the thermal expansion curves taken on warming and cooling. At this field we observe two transitions, the second order magnetic ordering transition near 12\,K and the sharp peaks at 7.1\,K (arrow). This peak corresponds to the I-VII phase boundary. It exhibits minor hysteresis (about 0.1\,K) and a very tiny relative length change (order 0.6\,ppm).

Both transitions also appear in the 1.4\,T heat capacity data (Fig.~\ref{fig:HighFieldPhases}\textbf{(c)}). Here again, we see first order-like peaks in $C_p(T)/T$ near 7.1\,K. We estimate the entropy of transformation to be 2.7(1)\,mJ/mol\,K. The ordering of Eu$^{2+}$ magnetic moments is expected to release $R\ln(8) = 17.3$\,J/mol\,K. This 7.1\,K transition corresponds to a minuscule 0.016\% of this, implying a subtle change between magnetic phases.

Despite careful examination, we are unable to identify a signature of this I-VII phase transition in DC-magnetization measurements. We do observe a subtle jump in the AC magnetic susceptibility. Figure~\ref{fig:HighFieldPhases}\textbf{(d)} presents $\chi'(T)$ at 1.4\,T taken on cooling. A small step is observed at 7.0\,K corresponding the peaks in $\alpha_a$ and heat capacity. This weak signature in $\chi'(T)$ implies a 0.6\% change in the slope of $M(H)$. This transition clearly has only a small affect on the net magnetization of the material and explains why previous studies have not identified it. 

The data presented in Fig.~\ref{fig:HighFieldPhases} reveals a subtle transition near 7\,K that looks like a first order phase transition. Next we will describe why we believe this is a real transition. First, we observe this transition in three independent measurements on the same crystal and the transition temperatures are in excellent agreement. This indicates they likely have the same origin. In fact, we observed this step feature in $\chi'(T)$ in Fig.~\ref{fig:HighFieldPhases}\textbf{(d)} on a second crystal at the same temperature. Finally, $\chi''(T)$ is slightly elevated within phase VII indicating increased damping.

We do not believe the I-VII transition is related to an impurity phase or miss-aligned grains. Minor impurities are unlikely to produce features in heat capacity or dilatometry as these average over the entire sample volume. Also, the crystal used for this study was visibly a single grain and surface phases were removed when the crystal was shaped for dilatometry (see section~\ref{sec:Methods_Measurements}).

Our final argument that phase VII is present in the $\bm{H}\parallel [001]$ phase diagram of EuAl$_{4}$ is based on the relationship between the phase boundaries. If the I-VII transition arose from an impurity phase or miss-aligned grain we would expect to see signatures extending beyond the previously observed transitions. Instead, we clearly see that the transition only appears in Fig.~\ref{fig:HighFieldPhases}\textbf{(a)} for fields between the phase VI and the field-polarized phase (above 1.8\,T). 

We have now laid out our evidence for a new phase in EuAl$_{4}$ and we will discuss the implications of phase VII in section~\ref{sec:Discussion_NaturePhaseVII}. 

\subsection{Magnetic phase diagram}
\label{sec:Results_PhaseDiagram}

\begin{figure*}
	\includegraphics{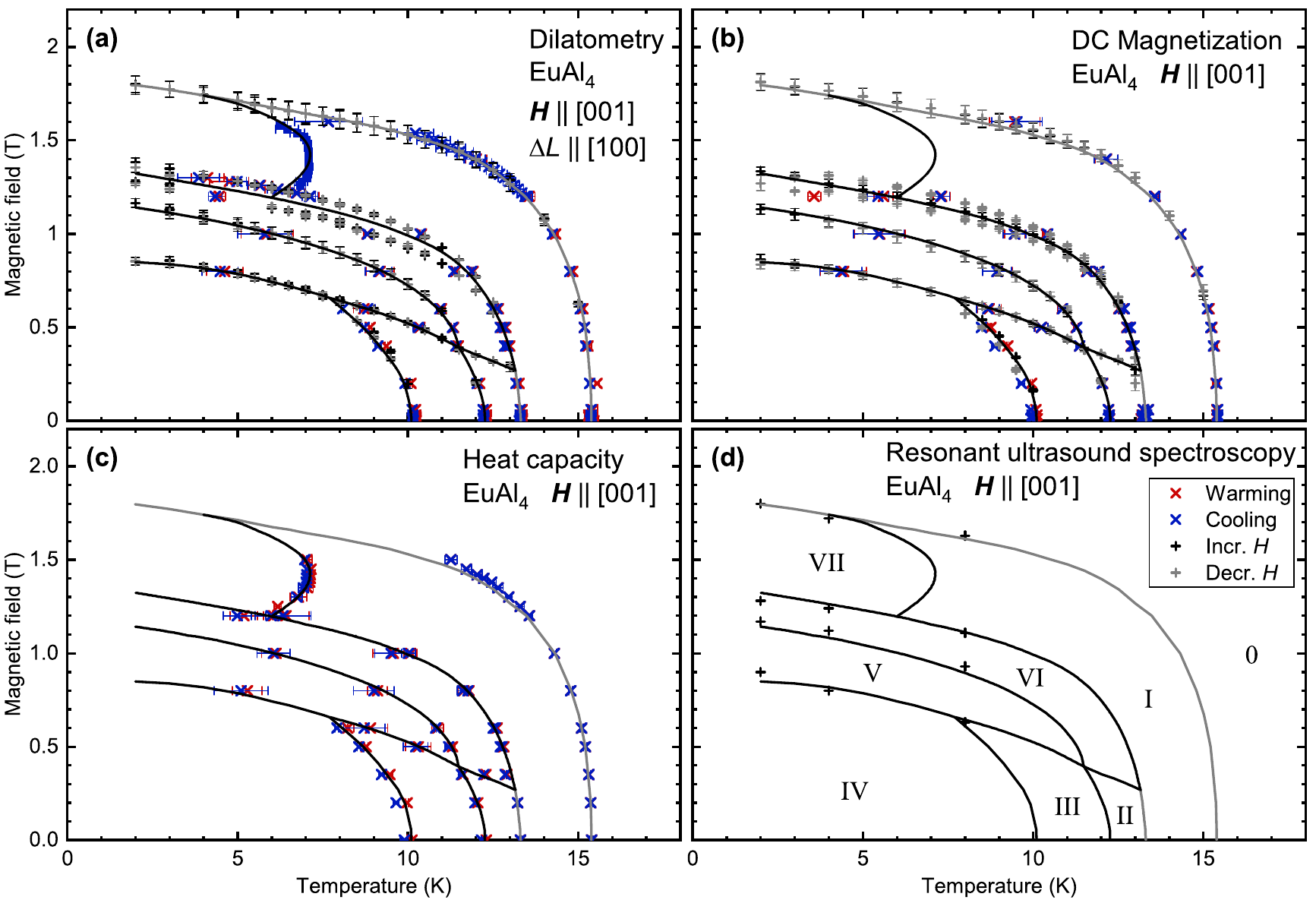}
	\caption{\label{fig:PhaseDiagramCompare} Comparison of transitions observed in each measurement. Solid lines represent the phase boundaries we propose and the phases are labeled with roman numerals in panel \textbf{(d)}. Black and gray lines represent transitions with first and second order character. For every panel, $\times$ symbols indicate transitions observed in temperature dependent measurements with red and blue representing warming and cooling, respectively. $+$ signs indicate transitions identified in field dependent measurements with black and gray symbols for increasing and decreasing field, respectively. Error bars represent estimated half-widths of the observed transitions. Magnetic field is the applied field, not corrected for demagnetization.
	}
\end{figure*}

So far, we have presented the evidence for individual phase transitions based on our thermodynamic measurements. Now, we summarize the transitions identified in our data in Fig.~\ref{fig:PhaseDiagramCompare}. The $\times$ and $+$ symbols represent transitions identified in temperature-dependent and field-dependent measurements respectively. Error bars represent estimated transition widths. Solid lines trace the phase boundaries identified in the data.

There is exquisite agreement between the observed transitions extracted for the dilatometry, magnetization and heat capacity measurements in panels \textbf{(a)}, \textbf{(b)}, and \textbf{(c)} of Fig.~\ref{fig:PhaseDiagramCompare}. We also see that transitions identified in temperature-dependent ($\times$) and field-dependent ($+$) data sets track the same phase boundaries. Critically, the first three measurements were done on the same crystal. This eliminates the challenge of comparing measurements on samples with different demagnetization factors (see Appendix \ref{sec:Appendix_Demag}). 

Resonant ultrasound spectroscopy (RUS) also shows evidence for the low temperature magnetic transitions (see Appendix \ref{sec:Appendix_RUS}). The critical fields we identified in field-dependent measurements are plotted in Fig.~\ref{fig:PhaseDiagramCompare}\textbf{(d)} and show some agreement with the phase boundaries derived from the other three measurements. A different demagnetization factor for the RUS sample and temperature stability issues could explain the differences. All together, our thermodynamic results provide strong evidence for the phase diagram we propose.

\section{Discussion}
\label{sec:Discussion}

Now, we will explore a few key aspects of our data. First, in section \ref{sec:Discussion_TetragonalPhases} we will discuss which phases show strong distortions from tetragonal symmetry. Then we will explore the thermodynamic character of the magnetic phases and transitions in section \ref{sec:Discussion_Phases+Transition}. Next, in section \ref{sec:Discussion_NaturePhaseVII}, we discuss the nature of phase VII where a topological Hall effect has been claimed. Finally, we will discuss implications of our results and outlook for the magnetic phases of EuAl$_{4}$ in section \ref{sec:Discussion_EuAl4Outlook}.

\subsection{Distortions from tetragonal symmetry}
\label{sec:Discussion_TetragonalPhases}

\begin{figure*}
	\includegraphics{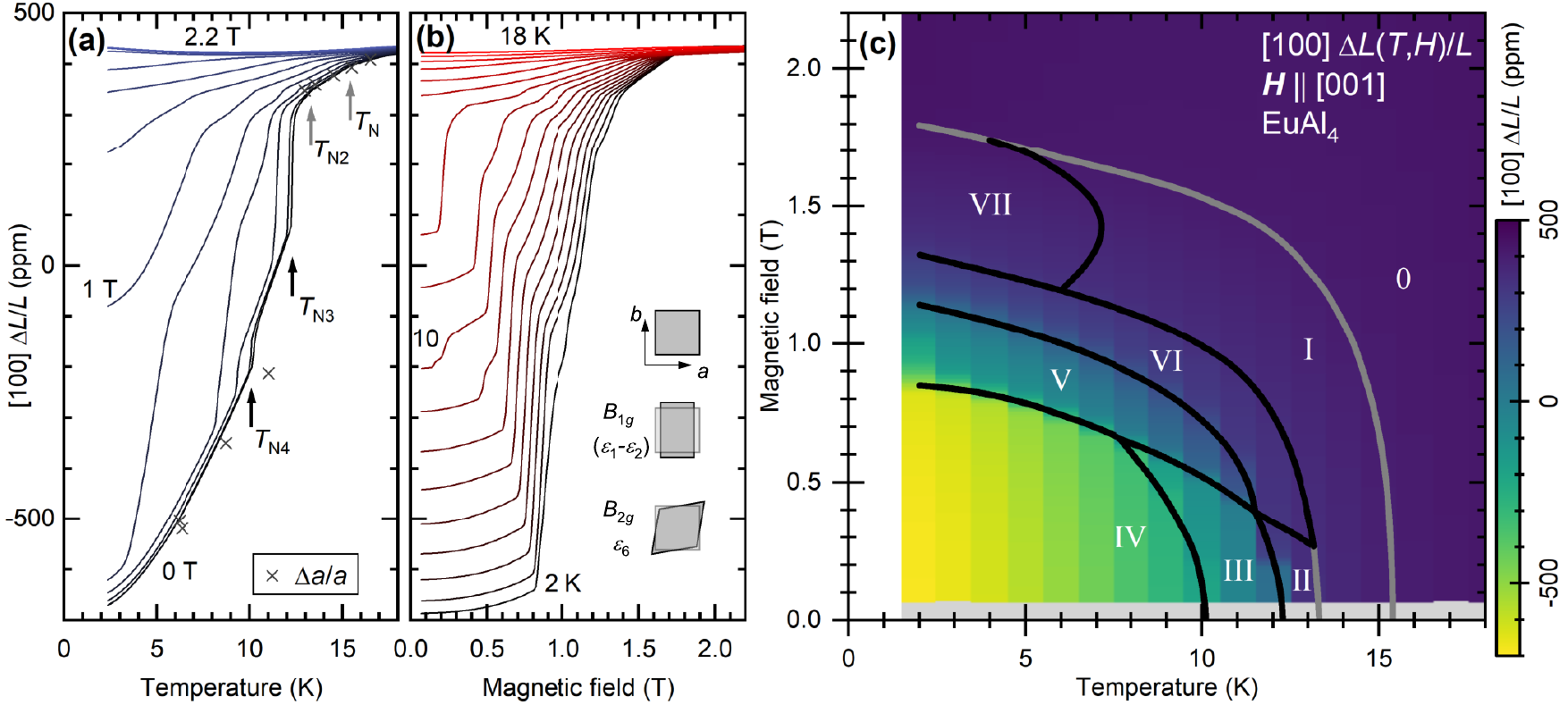}
	\caption{\label{fig:TetragonalPhases} Dilatometry results illustrating which phases show significant length changes. Panel \textbf{(a)} shows the dilation vs temperature behavior taken on warming at fields from 0 to 2.2\,T in 0.2\,T steps. The sample is consistently longer at each higher applied field. Gray $\times$'s show the relative change in the $a$ lattice parameter pulled from Shimomura et al.\cite{Shimomura2019_CDW-EuAl4}. The close agreement with their results an our 0\,T thermal expansion curve suggest that the sample is nearly de-twinned (see sec.~\ref{sec:Discussion_TetragonalPhases}). Panel \textbf{(b)} presents the dilation vs field behavior taken on increasing field with 1\,K steps. The step-like and kink-like transitions evolve to lower fields at each higher temperature. The inset shows the in-plane distortion modes of a tetragonal system. Our [100] dilatometry measurement is sensitive to the $B_{1g}$ mode. \textbf{(c)} color map representing the relative length of the EuAl$_{4}$ crystal along [001] with the phase diagram superimposed. These measurements were taken at constant temperatures between 2 and 18\,K with increasing field. Dark colors represent values closer to the paramagnetic phase 0 (i.e. 16\,K, 0\,T). Brighter colors represent $H,T$ values were the crystal is significantly shorter.
	}
\end{figure*}

In this section we will discuss which magnetic phases in the $\bm{H}$\,$\parallel$\,$[001]$ EuAl$_{4}$ phase diagram show strong deviations from tetragonal symmetry. Multi-$Q$ topological spin textures can be enabled by tetragonal symmetries. Large lattice distortions break these symmetries and favor single-$Q$, non-topological orders. Therefore, it is important to examine which low-temperature magnetic phases in EuAl$_{4}$ exhibit strong symmetry breaking strains. Phases III and IV have already been demonstrated to have lower symmetry\cite{Shimomura2019_CDW-EuAl4} based on an $a \ne b$ distortion observed by x-ray diffraction (a $B_{1g}$ mode). Our dilatometry results suggest that phase V exhibits the same distortion from tetragonal.

Figure \ref{fig:TetragonalPhases} presents relative length changes of EuAl$_{4}$ along [100] as a function of field and temperature. Importantly, this measurement reveals strong breaking of $a = b$ tetragonal symmetry by a $B_{1g}$ distortion of the unit cell in phases III, IV, and V. In Voigt notation this is a $(\epsilon_1-\epsilon_2)$ strain of the high-temperature $4/mmm$ structure and is depicted as an inset in Fig.~\ref{fig:TetragonalPhases}\textbf{(b)}. 

Panel \ref{fig:TetragonalPhases}\textbf{(a)} presents the relative length change of the crystals on warming at constant field. First consider the lowest, zero-field curve. Note the large magnitude of the length change we observe. At zero field, the sample changes in length by an impressive 1100\,ppm (0.11\%) between 2 and 20\,K suggesting a strong coupling of Eu magnetic order to the lattice. For comparison, we measured a 6.4\,ppm thermal expansion of copper over the same temperature range.

Critically, the large length changes we observe in our dilatometry data provides a good estimate of the change in size of the unit cell and the magnitude of $B_{1g}$ strain. Consider the gray $\times$'s in Fig.~\ref{fig:TetragonalPhases}\textbf{(a)}. These represent the relative change of the $a$ lattice parameter determined by x-ray diffraction in Shimomura et al.\cite{Shimomura2019_CDW-EuAl4} (shifted to match the dilatometry data at 16.5\,K). To our surprise, the zero-field dilation data tracks the lattice parameter change closely. This indicates that temperature dependence of $\Delta L/L$ presented in Fig.~\ref{fig:TetragonalPhases} is largely driven by change of $\Delta a/a$ brought about by $B_{1g}$ distortion observed by Shimomura et al. in phases III and IV. 

It is important to note that first order transitions and the thermal expansion of phases III and IV make significant contributions to the length change. At zero field, the sample experiences abrupt expansions of 230 and 71\,ppm on warming through $T_{\mathrm{N3}}$ and $T_{\mathrm{N4}}$, respectively. Curiously, the thermal expansion within phases III and IV are even more dramatic, lengthening by 204 and 472\,ppm, respectively. This suggests that evolution of the magnetic order within these phases drives a significant fraction of the symmetry breaking strain we observe. 

Now, examine the length changes in phases I and II above $T_{\mathrm{N3}}$. These phases account for a modest part of the low-temperature length changes (50 and 37\,ppm, respectively). This suggest that there is no significant $B_{1g}$ distortion. It is important to note that our [100] dilatometry measurement is not sensitive to a $B_{2g}$ in-plane shear distortion ($\epsilon_{6}$) depicted in the inset of Fig.~\ref{fig:TetragonalPhases}\textbf{(b)}.

Shimomura et al. state that phases I and II appear to be tetragonal\cite{Shimomura2019_CDW-EuAl4} because they did not observe peak splitting indicating in-plane distortions. Despite this, a new diffraction report reveals that the CDW itself breaks tetragonal symmetry\cite{Ramakrishnan2022_OrthorhombicCDW-EuAl4} at temperatures well above the magnetic transitions. Their data\cite{Ramakrishnan2022_OrthorhombicCDW-EuAl4} is best modeled by an orthorhombic structure that would allow a $B_{2g}$ strain. Critically, they also do not see a resolvable splitting of diffraction peaks indicating that the distortion of the unit cell is limited. Therefore, phases I and II have only weakly broken tetragonal symmetry and might still be able to host multi-Q magnetic order. 

In summary, our zero-field dilatometry data are consistent with phases I and II being effectively tetragonal while phases III and IV reveal a strong deviation from tetragonal symmetry. Now that we have explored the zero-field thermal expansion results, we will examine the in-field behavior. The curves in Fig.~\ref{fig:TetragonalPhases}\textbf{(a)} present the thermal expansion curves at fields from 0 to 2.2\,T in 0.2\,T steps. We immediately observe two points. Sample length monotonically increases with field and we can see evolution of the step and kink phase transitions as a function of field. In other words, applying field and increasing temperature favors phases with smaller $B_{1g}$ strains.

Figure \ref{fig:TetragonalPhases}\textbf{(b)} reveals both of the effects more clearly. This plot depicts sample deformation with increasing magnetic field at constant temperatures from 2 to 18\,K in 1\,K steps. Once again, we observe that the crystal lengthens along [100] as we increase both field and temperature. We can clearly observe many step-like and kink-like transitions evolving with temperature.

Our dilatometry results are summarized in Figure~\ref{fig:TetragonalPhases}\textbf{(c)} highlighting which phases show a strong distortion from tetragonal symmetry. This plot presents the relative length change of the EuAl$_{4}$ crystal along [100] as a function of temperature and applied field along the $c$-axis. Colors represent the value of $\Delta L/L$. Darker shades represent $T, H$ values where the sample is longer and close to its length in the paramagnetic phase 0 above $T_{\mathrm{N}}$. Lighter colors represent regions where the crystal is significantly shorter. These data were all taken at constant temperature with increasing fields. The superimposed phase diagram labels the phase fields we have determined.

In Fig.~\ref{fig:TetragonalPhases}\textbf{(c)}, phases III, IV, and V have lighter colors representing a large contraction from the the paramagnetic phase 0. This implies that these three phases possess a strong $B_{1g}$ distortion from tetragonal symmetry. In contrast, the sample length in phases I, II, VI, and VII are quite close to the length of phase 0, represented by dark colors. This indicates that these phases likely have $a = b$ and only weak orthorhombic distortions, at best.

This leads us to the implications of our dilatometry results for skyrmions and merons in EuAl$_{4}$. The strong $B_{1g}$ distortion we observe in phases III, IV and V make them unlikely hosts for the multi-$Q$ topological magnetic textures. Our dilatometry results observe only weak length changes along [100] in phases I, II, VI, and VII suggesting that they have nearly equal $a$ and $b$ lattice parameters. It is unlikely that these phases have full tetragonal symmetry because of the orthorhombic CDW modulation\cite{Ramakrishnan2022_OrthorhombicCDW-EuAl4}. Despite this, previous diffraction reports\cite{Shimomura2019_CDW-EuAl4,Ramakrishnan2022_OrthorhombicCDW-EuAl4} and our dilatometry results suggest that there may only be weak breaking of tetragonal symmetries. Therefore, phases I, II, VI, and VII might host multi-Q phases such as skyrmion or meron crystals.

\subsection{Characteristics of phases and transitions}
\label{sec:Discussion_Phases+Transition}

Next, we will explore some characteristics of the magnetic phases and their transitions. Our thermodynamic measurements provide insights and raise new questions about these phases. We will explore what differentiates phases I and II as well as III and IV. Then we will take a look at the character of II-VI and III-V transitions we have identified. Then, we will briefly discuss the unusual nature of the I-VI transition.

\subsubsection{Phases I and II}
\label{sec:Discussion_I-II-transition}

First we will take a closer look at phases I and II and the transition between them. Kaneko et al.\cite{Kaneko2021_EuAl4-SCNeutronDiffraction} report magnetic propagation vectors of $\bm{Q} = (q,q,0)$ and $(q,-q,0)$\,r.l.u. in both phases (see Fig.~\ref{fig:IntroFigure}). The magnetic periodicity within these phases are nearly equal with $q =$ 0.086(4) and 0.085(4)~r.l.u. at 13.5 and 12.5\,K, respectively. We suspect these phases are almost tetragonal and share nearly equal $\bm{Q}$'s. So, what differentiates these phases?

Our magnetic measurements offer one clue to the differences between phases I and II. Phases I and II have different low-field magnetic behavior despite their common magnetic propagation vectors. The transition between them appears first order in dilatometry, heat capacity and AC-susceptibility, but only for small applied field. For fields larger 5\,mT the transition appears second order. This might be due to weak ferromagnetism in phase II.

Note that phase II has different behavior in low field (0.01\,T) DC magnetization measurement and AC-susceptibility data (Figs.~\ref{fig:ZeroFieldPhases}\textbf{(d)} and \textbf{(e)}, respectively). On cooling through $T_{\mathrm{N2}}$ into phase II the DC magnetization increases abruptly into a plateau with a negative slope. In contrast, there is a peak in the AC-susceptibility, $\chi'$ at the transition and phase II has a lower susceptibility with a rising slope. Finally, the out-of-phase AC-susceptibility plot, $\chi''(T)$, in Fig.~\ref{fig:ZeroFieldPhases}\textbf{(e)} has a peak at $T_{\mathrm{N2}}$ and a plateau within phase II, 80\% higher than in the adjacent phases. This indicates that there is lossy mechanism within the phase. 

Together, these magnetic measurements suggest that phase II has a small ferromagnetic contribution. This would explain why the DC moment at 10\,mT increase from phase I to II and the AC-susceptibility is smaller with its 0.5\,mT excitation field. Low field magnetic hysteresis would also produce the more lossy $\chi''$ signal we observe. Finally, the critical field for this hysteresis may be quite small because an AC-susceptibility measurement in 5\,mT no longer shows a plateau in $\chi''(T)$.

These phases have some subtle difference in details of the magnetism. First, $T_{\mathrm{N2}}$ represents the appearance of a weak ferromagnetic component on cooling into phase II. This would explain the plateau in $M(T)$ in Fig.~\ref{fig:ZeroFieldPhases}\textbf{(d)} and zero-field magnetic loss (Fig.~\ref{fig:ZeroFieldPhases}\textbf{(e)}). Second, the transition might represent a change from single-$Q$ to two-$Q$ version of the ($q,q,0$) magnetic order. In this case, we might observe an extension of the unit cell along [110] (a $\epsilon_{6}$ shear) in the single-$\bm{Q}$ phase. We are not sensitive to this distortion with our [100] dilatometry measurement. Resonant x-ray or neutron diffraction might be able to observe peaks at harmonics of the primary magnetic wave vectors such as ($2 q,0,0$).\cite{Khanh2020_NanometricSkyrmionLatticeGdRu2Si2} Finally, $T_{\mathrm{N2}}$ may signal the emergence of another magnetic modulation component but, no additional peaks were reported in phase II with neutron diffraction\cite{Kaneko2021_EuAl4-SCNeutronDiffraction}.

Phases I and II deserve a bit more attention in future investigations. They appear to show common magnetic wave vectors but have distinctly different low field magnetic behavior. We propose that x-ray diffraction measurements could help identify these phases. Single-$Q$ magnetism couples to a [110] extension strain that should be evident in high-resolution x-ray diffraction measurements. Resonant x-ray diffraction might also identify the higher order magnetic peaks indicating two-$Q$ magnetic textures.

\subsubsection{Phases III and IV}
\label{sec:Discussion_III-IV-transition}

In this section, we will discuss the characteristics of the two low-temperature and low-field phases, III and IV. These phases both clearly break tetragonal symmetry. Diffraction results reveal distinctly different $a$ and $b$ lattice parameters\cite{Shimomura2019_CDW-EuAl4} reflected in our dilatometry results (see sec.~\ref{sec:Discussion_TetragonalPhases}). The dilatometry results also show that the phases have similar dramatic thermal expansion behavior ($\alpha_a$'s about 100\,ppm/K in Fig.~\ref{fig:ZeroFieldPhases}). Finally, these phases share similarities in their magnetic order. Neutron diffraction reveals that phases III and IV host incommensurate antiferromagnetism with wave vectors ($q,0,0$) and ($0,q,0$). \cite{Kaneko2021_EuAl4-SCNeutronDiffraction}. These similarities between phases III and IV raise two questions: What distinguishes these two phases and why are they separated by a first order, hysteretic transition at $T_{\mathrm{N4}}$? 

The III-IV transition at $T_{\mathrm{N4}}$ might herald the appearance of an additional component of the ($q,0,0$) modulated magnetism. It is curious that this would happen abruptly at a hysteretic first order transition.  Alternatively, the transition might represent a discontinuity in $q(T)$. To this point, distinctly different magnetic wave-vectors are observed at 11.5 and 4.3\,K with $q$ = 0.17(1) and 0.194(5)\,r.l.u., respectively\cite{Kaneko2021_EuAl4-SCNeutronDiffraction}. Finally, it is especially surprising to find a moderate thermal hysteresis at $T_{\mathrm{N4}}$ as our dilatometry results suggest that the sample is nearly detwinned (sec.~\ref{sec:Discussion_TetragonalPhases}). 

Maybe the transition is driven to be first order by competition between the ($q,0,0$) magnetism and the charge density wave degree of freedom. X-ray diffraction results reveal that the lattice has a strong response on cooling into phase III \cite{Shimomura2019_CDW-EuAl4}. The intensity of the CDW satellite peaks drops by 21\% across $T_{\mathrm{N3}}$ and the intensity of the (600) peak falls by 33\%. The magnetic transitions at $T_{\mathrm{N}}$ and $T_{\mathrm{N2}}$ do not demonstrate such strong coupling to the lattice. The transition between phases III and IV might be a consequence of the competition between the magnetic and lattice modulations.

\subsubsection{II-VI and III-V transitions}
\label{sec:Discussion_II-VI_III-V-transitions}

In section~\ref{sec:Results_LowFieldTrans} we presented the evidence for phase transitions between phases II and VI as well phases III and V. Previous studies did not observe these transitions\cite{Shang2021_AHE-EuAl4,Nakamura2015_Transport+MagPropEuAl4+EuGa4}. This is likely because they used wider temperature and field steps. In this section we will discuss the nature of these transitions and their implications. 

The transition between phases II and VI appears clearly in both field-dependent and temperature-dependent measurements. At 12.5\,K the transition manifests as a step in the $M(H)$ curve in Fig.~\ref{fig:LowFieldPhasesVsH}\textbf{(d)}, but phases II and VI have similar slopes ($\frac{d M}{d (\mu_0 H)} = 3$\,$\mu_\mathrm{B}$/Eu T). Dilatometry measurements reveal that both phases show small length differences from the paramagnetic phase 0 (sec.~\ref{sec:Discussion_TetragonalPhases} and dark colors in Fig.~\ref{fig:TetragonalPhases}(a)) but they show different magneto-striction behavior. In Fig.~\ref{fig:LowFieldPhasesVsH}\textbf{(h)}, phase II has a magentostriction coefficient, $\lambda_a$, of 60-70\,ppm/T which falls to 30-40\,ppm/T in phase VI after a 4\,ppm expansion at the 0.32\,T transition. These changes in magnetic and dilation behavior suggest a relatively minor reconfiguration of the magnetic order.

In contrast, the III-V shows significant changes in the thermodynamic properties of EuAl$_{4}$. First, consider the dramatic change in magnetization at this transition (10.3\,K) observed in Fig.~\ref{fig:LowFieldPhasesVsTemp}\textbf{(d)}. This shows that phase V has a significantly higher magnetization than phase III. This is directly reflected in the $M(\mu_0 H)$ plots in Figs.~\ref{fig:LowFieldPhasesVsH}\textbf{(b)} and \textbf{(c)} as sharp jumps in the magnetization at 0.59 and 0.48\,T, respectively. The dilatometry also reveals significant sample expansion across this transition in panels \textbf{(f)} and \textbf{(g)}. Both $M(\mu_0 H)$ and $\Delta L(\mu_0 H)/L$ are notably non-linear in phase V with rising slopes with increasing field.

The transition between phases III and V represents a more dramatic change in the magnetic order than the II-VI transition. This transition is relatively broad in temperature and field-dependent measurements which might reflect demagnetization effects (see Appendix \ref{sec:Appendix_Demag}). The changes in magnetization and magnetostriction indicate that phases III and V respond to field in different ways. This suggests that the configuration of the magnetic order has changed dramatically. The jump in $M(\mu_0 H)$ might reflect a change from cycloidal or helical order in phase III to a fan or conical phase V. This could also explain the non-linear field dependence of properties of phase V in Figs.~\ref{fig:LowFieldPhasesVsH}\textbf{(b)}, \textbf{(c)}, \textbf{(f)}, and \textbf{(g)}. 

The II-VI and III-V transitions we report are clear additions to the $\bm{H}$\,$\parallel$\,$[001]$ EuAl$_{4}$ phase diagram. Phases II and VI show similar magnetic and magnetostricitive characteristics but the distinction between phases III and V appear more dramatic. A detailed examination of the magnetic order by single crystal diffraction under field will provide insights into the intricate competition between the magnetic and lattice subsystems at play EuAl$_{4}$.

\subsubsection{I-VI transition}
\label{sec:Discussion_I-VI-transition}

We will make a few brief comments about the I-VI phase transition. This phase boundary has unique character in this system. The transition consistently appears as a split feature in magnetic, heat capacity and dilatometric measurements. This double feature is clearly visible in Figs.~\ref{fig:LowFieldPhasesVsH}\textbf{(b)}, \textbf{(c)}, \textbf{(f)}, and \textbf{(g)} (1.05\,T at 9\,K and 0.95\,T at 10.5\,K). The thin $\frac{d M}{d (\mu_0 H)}$ plots show a shoulder-peak feature and the $\lambda_a(\mu_0 H)$ shows a double peak feature separated by roughly 0.06\,T.

In figure \ref{fig:PhaseDiagramCompare} we attempt to capture this split transition by plotting the estimated centers of peaks in $C_{p}(T)$ and the derivatives of $\Delta L/L$ and $M$. These points form two lines on either side of the boundary between I and VI phase fields. Our confidence in the double feature is strengthened by two observations. First, we observe strong agreement between dilatometry, magnetization and heat capacity measurements on the position of these two trends. Second, both temperature-dependent and field-dependent measurements produces features following the same trend. 

The split I-VI phase transition might represent an unusual manifestation of a transition broadened by demagnetization effects (Appendix \ref{sec:Appendix_Demag}) or there might be an additional phase in this between these peaks. We suggest that future experiments devote some attention the regions between phases I and VI and between VI and VII to uncover further clues about this unusual transition.

\subsection{What do we know about phase VII?}
\label{sec:Discussion_NaturePhaseVII}

Finally, we will examine the nature of our most interesting new phase in EuAl$_{4}$, phase VII. We propose that this distinct phase appears below 7\,K between 1.2 and 1.8\,T based on our measurements presented in section \ref{sec:Results_HighFieldTrans}. Figure \ref{fig:HighFieldPhases} reveals a first order transition bounding this region based on dilatometry, heat capacity and AC magnetic susceptibility measurements.

Phase VII has special importance in this system as Shang et al. revealed that this $H,T$ region may host a topological Hall effect\cite{Shang2021_AHE-EuAl4}. As the authors discuss, this signal does not appear to arise from an anomalous Hall contribution due to the uniform magnetization. Our data is in agreement, as we do not observe a significant change in the DC magnetization across the I-VII phase transition that would yield a change in the anomalous contribution. This suggests that there is a topological Hall contribution generated by a non-collinear magnetic texture in phase VII.

Although skyrmions are implicated in Ref.~\cite{Shang2021_AHE-EuAl4}, other topological magnetic phases could generate the observed Hall contribution in this region\cite{Wang2021_TetCentrosymmMeron+Skyrmion+VortexCrystals}. For example, some meron-crystals host a net vector spin chirality which generates the topological Hall contribution.

So, does EuAl$_{4}$ host a skyrmion or meron crystal in phase VII? In short, maybe. These magnetic orders could explain the pocket-like phase field. First, all the meron and skyrmion crystals are consistently bounded by first order phase transition\cite{Wang2021_TetCentrosymmMeron+Skyrmion+VortexCrystals,Khanh2020_NanometricSkyrmionLatticeGdRu2Si2,Spachmann2021_MagnetoelasticCouplingGd2PdSi3}. This appears to be the case for phase VII (see, Fig.~\ref{fig:HighFieldPhases}\textbf{(a)}). 

Second, the skyrmion crystal and one of the meron crystals in Wang et al. that have a finite vector spin chirality (SkX and MX-I) appear for finite applied field and have net magnetization. In fact, for the parameters they explore, the skyrmion crystal phase is generally adjacent to or near the field polarized phase 0. This is exactly the case for phase VII in EuAl$_{4}$. It appears at low temperature just below the saturation field with nearly saturated magnetization. This observation is suggestive of a skyrmion crystal in phase VII.

Next, we return to broken tetragonal symmetry in the magnetic phases of EuAl$_{4}$. The multi-$Q$ meron crystals and a square skyrmion should be hosted by tetragonal materials. Domains of the 3-$Q$ triangular skyrmion crystal discussed in Ref.~\cite{Wang2021_TetCentrosymmMeron+Skyrmion+VortexCrystals} might only lead to weak strains of a bulk sample. Our results suggest that both phases I and VII are are nearly tetragonal based on our dilatometry measurements but the coexisting CDW modulation appears to break 4-fold symmetry\cite{Ramakrishnan2022_OrthorhombicCDW-EuAl4}. This situation would favor single-$Q$ magnetic orders. It is possible that a weakly modified tetragonal phase could host the multi-$Q$ magnetic textures and produce a topological Hall signal in phase VII.

An argument against skyrmions in phase VII is the $M(\mu_0 H)$ behavior we observe (see Fig.~\ref{fig:MH}). Like others before, we observe a simple change in slope at the saturation field \cite{Nakamura2015_Transport+MagPropEuAl4+EuGa4,Shang2021_AHE-EuAl4}. In many experimental reports\cite{Spachmann2021_MagnetoelasticCouplingGd2PdSi3,Khanh2020_NanometricSkyrmionLatticeGdRu2Si2} and theoretical predictions\cite{Leonov2015_MultiplyPeriodicState+SkyrmionsFrustratedMagnets,Hayami2016_Bubble+SkyrmionCrystalsFrustratedEasyAxisMagnets,Yambe2021_SkyrmionCentrosymmetricItinerantMagnetsWOMirror}, skyrmion crystals generate a terrace-like feature in $M(\mu_0 H)$ with reduced slope. Our $\chi'(T)$ data shows a very weak reduction in $dM/dH$ on cooling into phase VII but no clear terrace region is obvious in $M(\mu_0 H)$ curves. In addition, we would not expect a skyrmion or meron crystal to exist with a nearly saturated magnetization.

\subsection{Outlook}
\label{sec:Discussion_EuAl4Outlook}

Finally, we will examine experimental tests for skyrmion/meron phases in EuAl$_{4}$. Phase VII should be checked for these topological magnetic textures. Three techniques are commonly employed to identify these magnetic configuration and differentiate them from their mundane, single-$Q$ counterparts. Real-spacing imagining of skyrmions is often performed using Lorentz transmission electron microscopy. Unfortunately, phase VII appears below 7\,K, a challenging temperature regime for cold stages. Single crystal neutron or resonant x-ray diffraction under a [001] magnetic field are both clear options. We would expect to observe harmonics of the magnetic wave-vectors in the case of mulit-$Q$ order (e.g. intensity at $(2q,0,0)$ for $(q,q,0)$+$(q,-q,0)$ order). Neutron diffraction, although clearly possible\cite{Kaneko2021_EuAl4-SCNeutronDiffraction}, is challenged by Eu's strong neutron absorption. This makes resonant x-ray scattering a strong choice. 

To summarize, skyrmion or meron magnetic textures are good candidates for producing the Hall signal observed EuAl$_{4}$ within phase VII. The characteristics of this phase and its location in the phase diagram are reminiscent of a skrmion or meron crystal. We propose an in-field resonant x-ray diffraction experiment to examine phase VII for evidence of these topological magnetic textures. In addition, we believe that a diffraction study of the other $\bm{H}$\,$\parallel$\,$[001]$ magnetic phases will help uncover the competing interactions that drive the complex magnetism in EuAl$_{4}$.

\section{Conclusions}
\label{sec:Conclusions}

EuAl$_{4}$ shows a rich variety of low-temperature orders. It hosts both lattice modulations and complex magnetic phases. We explored the $\bm{H}$\,$\parallel$\,$[001]$ phase diagram in great detail using a series of thermodynamic measurements. Dilatometry, heat capacity, DC magnetization, AC magnetic susceptibility, and resonant ultrasound spectroscopy revealed not only the numerous transitions but also clues about their nature.

We observe all the previously reported low-temperature transitions including $T_{\mathrm{CDW}}$ and magnetic transitions. In addition, we also report three new phase transitions giving a total of seven magnetic phases under a $c$-axis magnetic field. The resulting detailed magnetic phase diagram (Fig.~\ref{fig:CartoonPhaseDiagram}) is the first important outcome to this study and will guide future investigations of the numerous magnetic phases of EuAl$_{4}$.

We discussed the characteristics of these phases and suggest that the distinction between phases be examined in detail. In particular, the I-II and III-IV transitions have curious features and deserve further study. It is not clear what differentiates the similar phases. 

Finally, we present evidence of a consequential new phase, phase VII. Its phase field directly coincidences with the region were a skyrmion phase was proposed based on an enhanced anomalous Hall signal. Our measurements suggest that this phase does have some characteristics of topological magnetic texture. We advocate for a detailed diffraction study to examine this potential skyrmion phase and look for distortions from tetragonal symmetry. In addition, determining the order parameters that describe the other phases in the system will offer insights into the competing interactions which underlie the intricate magnetic phase diagram of EuAl$_{4}$ we delineate in this study.

\begin{acknowledgments}
	\label{sec:Acknowledgment}
	We would like to thank Christian Batista, Shang Gao, Seunghwan Do, and Andrew Christianson for their helpful discussions about topological magnetic textures. We would like to thank Brian Chakoumakous for his low temperature single crystal XRD measurements. M.Y. Hu and E. E. Alp (Argonne) are acknowledged for assistance for NRIXS data acquisition at 3-ID.
	
	Research was supported by the U. S. Department of Energy, Office of Science, Basic Energy Sciences, Materials Sciences and Engineering Division (under contract number DE-AC05-00OR22725). JRT was supported by ORNL Laboratory Directed Research and Development (LDRD) funds. WRM acknowledges partial support for writing from the Gordon and Betty Moore Foundation’s EPiQS Initiative, Grant GBMF9069. This research used resources of the Advanced Photon Source, a U.S. Department of Energy (DOE) Office of Science User Facility operated for the DOE Office of Science by Argonne National Laboratory under Contract No. DE-AC02-06CH11357.
	
\end{acknowledgments}


%

\appendix

\section{Resonant ultrasound spectroscopy}
\label{sec:Appendix_RUS}

\begin{figure}
	\includegraphics{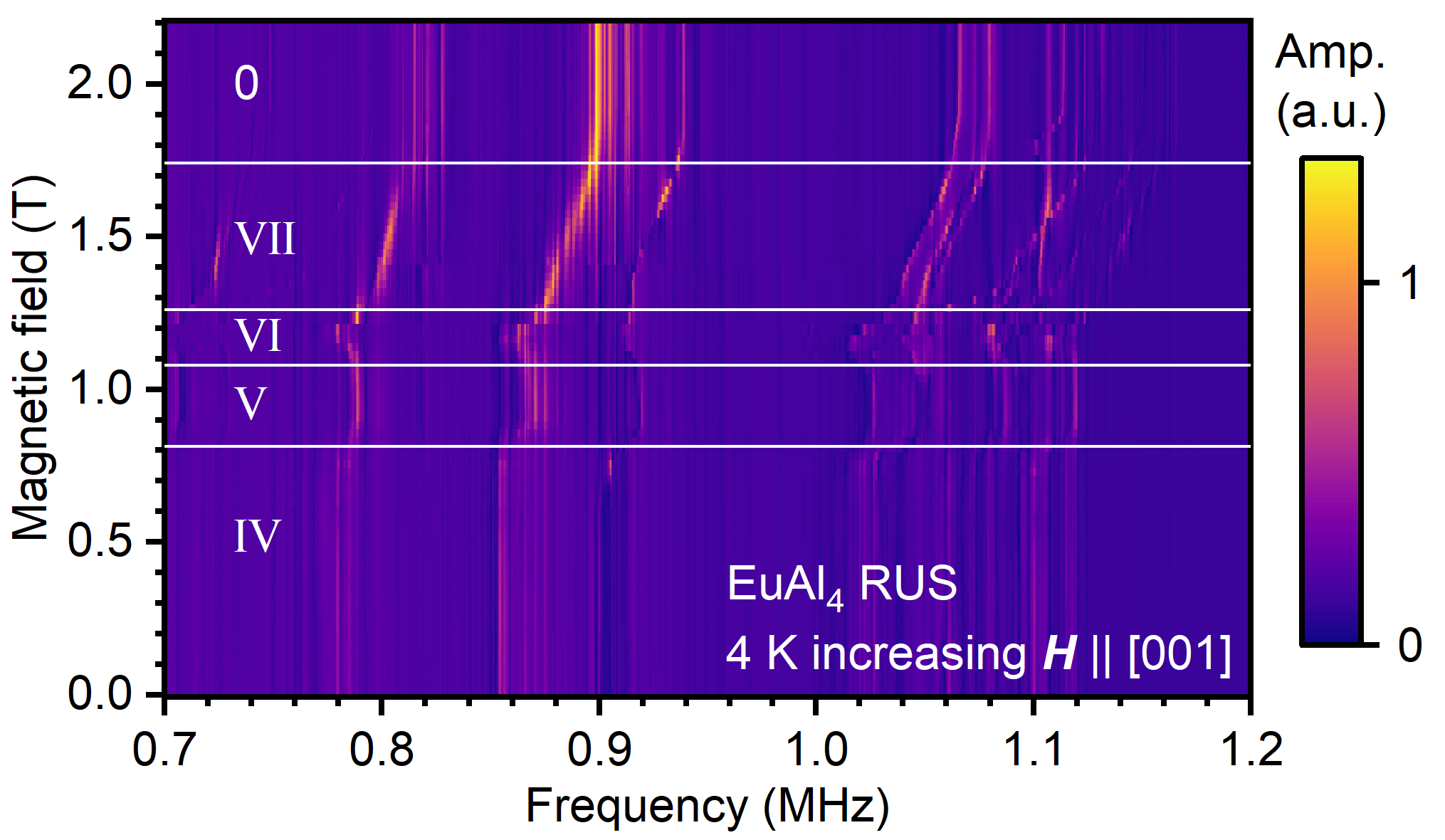}
	\caption{\label{fig:RUS-4K} Resonant ultrasound spectroscopy from the EuAl$_{4}$ sample under increasing $c$-axis magnetic field. Bright colors represent stronger mechanical resonant response of the sample near resonant modes. Each row represents a spectrum taken while ramping field. The horizontal lines shows the critical fields of the metamagnetic transitions based on the other measurements in this paper.
	}
\end{figure}

Low temperature field-dependent resonant ultrasound spectroscopy (RUS) shows clear signals of the magnetic transitions in EuAl$_{4}$. Figure \ref{fig:RUS-4K} presents resonant ultrasound spectra of a EuAl$_{4}$ sample on increasing field at 4\,K. Bright colors represent stronger mechanical response corresponding to resonant modes. Just as we discussed in section \ref{sec:Results_ChargeDensityWave}, shifts in the frequencies of resonant modes reflect changes in the elastic modulii. The four metamagnetic transitions (horizontal white lines) correspond to jumps in the peak positions or changes in the rate of frequency change vs field. On the whole, resonant modes shift to higher frequencies at higher field reflecting a stiffening of the lattice. Low-temperature resonant ultrasound spectroscopy (RUS) measurements were challenging to interpret and temperature dependent measurements suffered very large temperature errors. 

\section{Demagnetization affects}
\label{sec:Appendix_Demag}

\begin{figure}
	\includegraphics{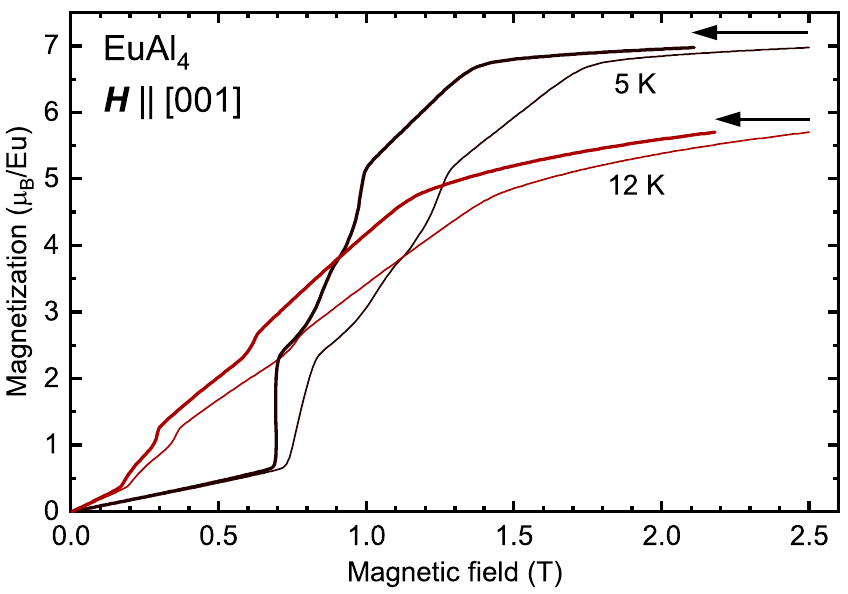}
	\caption{\label{fig:MHdemag} Magnetization vs field plots for EuAl$_{4}$ illustrating demagnetization corrections. The thin lines represent measured $M(\mu_0 H)$ plots. The thicker plots present $M(\mu_0 H_{\mathrm{internal}})$ based on an estimated demagnetization correction to the applied field. Arrows illustrate the mapping of $H_{\mathrm{applied}} \rightarrow H_{\mathrm{internal}}$.
	}
\end{figure}

A magnetized sample produces an inhomogeneous internal field referred to as the demagnetization field. As a result, the local field varies throughout sample with values below that of the applied field. Throughout this study we do not correct for the demagnetization field (but it can be done; Ref.~\cite{Spachmann2021_MagnetoelasticCouplingGd2PdSi3}). Careful understanding of sample geometry is critical but the correction can be accomplished for spheres, rectangular prisms and ellipsoids\cite{Chen2002_DemagFactorsRectangularPrisms+Ellipsoids}.

The key impact of the demagnetization correction is a remapping of all field dependent data to lower field values based on the magnetization and sharpening first order transitions. The expression in SI is 
\begin{equation}\label{equ:DemagEqu}
	H_{\mathrm{internal}} = H_{\mathrm{applied}} - N M
\end{equation}
where $N$ is the demagnetization factor\cite{Chen2002_DemagFactorsRectangularPrisms+Ellipsoids}. $M$ and $H$ values need to be in A/m. Examples of this remapping are presented for 5 and 12\,K $M(\mu_0 H)$ plots in Fig.~\ref{fig:MHdemag} with a demagnetization factor of $N = 0.52$. This value was chosen because it gives vertical $M(\mu_0 H)$ at the first order transitions \cite{Stryjewski1977_Metamagnetism}. Notably, it agrees well with the magnetometric demagnetization factor of $N = 0.48$ which we estimate from the tables in Ref.~\cite{Chen2002_DemagFactorsRectangularPrisms+Ellipsoids} using our sample dimensions (sec.~\ref{sec:Methods_Measurements}). 

The mapping illustrated in Fig.~\ref{fig:MHdemag} reduces fields by up to 20\% due to the significant magnetization from the large europium moments. These likely explains the differences in reported fields for the metamagnetic phase transitions in Refs.~\cite{Nakamura2015_Transport+MagPropEuAl4+EuGa4} and \cite{Shang2021_AHE-EuAl4}. For example, these report saturation fields for $\bm{H}$\,$\parallel$\,$[001]$ of 1.5 and 2.0\,T, respectively. After demagnetization correction, we estimate a saturation field of 1.4\,T at 2\,K. 

Stryjewski and Giordano (Ref.~\cite{Stryjewski1977_Metamagnetism}) note that demagnetization effects have important implications for experimental appearance of first-order metamagnetic transitions. The demagnetization field leads to coexistence of the low and high field phases over a range of applied fields leading to a broadened transition. In addition, phase coexistence could limit and obscure any hysteresis across the first order transition. Without demagnization field (like in a needle-like sample) we would expect discontinuous $M(\mu_0 H)$ curves at first order metatmagnetic transitions.

The slope of $\frac{d M}{d H} \propto 1/N$, where $H$ is the applied field and $N$ is the demagnetization factor\cite{Stryjewski1977_Metamagnetism}. This means that transitions with larger changes in magnetization, $\Delta M$ will exhibit wider transitions due to demagnetization effects: $\Delta H \propto N\,\Delta M$.

This effect is pronounced at this III-V transition because has a large $\Delta M$ and a shallow slope in the phase diagram. Like many of the first order transitions in this system, the III-V transition did not show hysteresis with field or temperature dependent measurements but shows a broad transition width. The transition is broader than others we observe, especially in temperature-dependent measurements (Fig.~\ref{fig:LowFieldPhasesVsTemp} near 10.3\,K). 

The shallow slope of III-V phase boundary line also explains why the transition appears unusually broad in the temperature-dependent measurements in Fig.~\ref{fig:LowFieldPhasesVsTemp}. We estimate full width at half max of the transition to be 0.8\,K in Fig.~\ref{fig:LowFieldPhasesVsTemp} and 0.05\,T in the 10.5\,K field-dependent measurements (\ref{fig:LowFieldPhasesVsH}\textbf{(c)} and \textbf{(g)}). The ratio of these values is 0.06\,T/K very close to the absolute value of the slope of the III-V phase boundary line, 0.08\,T/K. This means that the anomalously broad of the transition in Fig.~\ref{fig:LowFieldPhasesVsTemp} is explained by the slope of transition line and demagnetization effects. Although not shown here, the transition between phases IV and V is also notably broad and has a large $\Delta M$.


\section{Magnetization vs Field}
\label{sec:Appendix_MH}

\begin{figure}
	\includegraphics{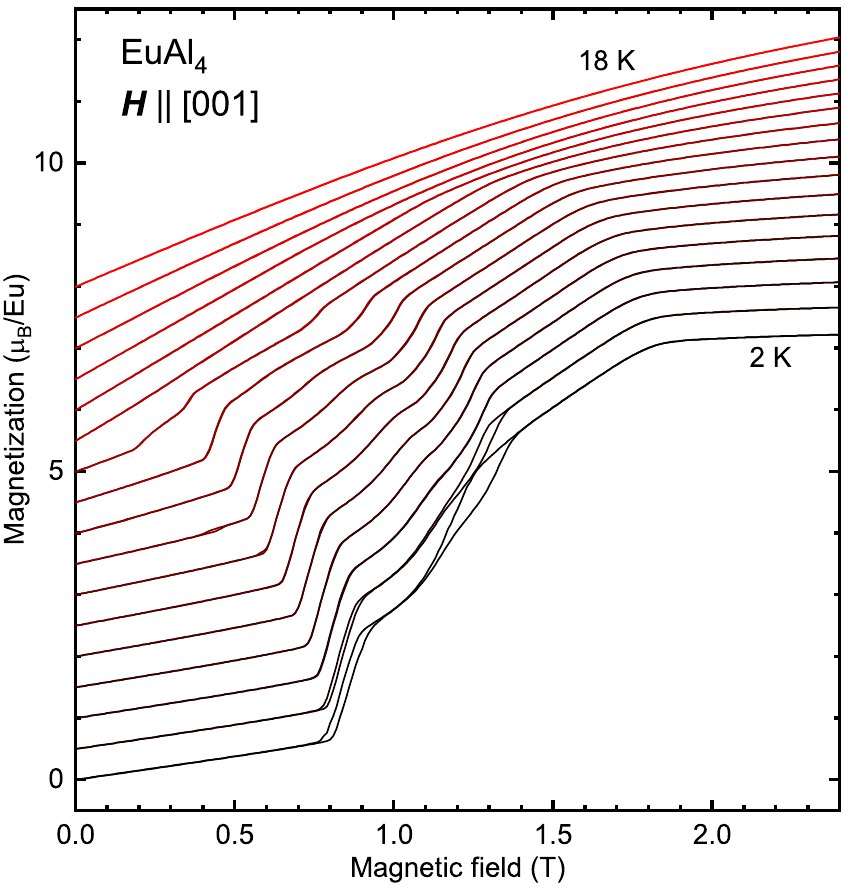}
	\caption{\label{fig:MH} $M(\mu_0 H)$ plots for EuAl$_{4}$ with field along the $c$-axis. Each curve presents the magnetization between 2 and 18\,K in 1\,K steps offset by 0.5\,\textmu$\textrm{B}$/Eu. Data obtained while increasing and decreasing field are presented for each temperature and hysteresis is evident at transitions in the 2 and 3\,K plots.
	}
\end{figure}

Figure \ref{fig:MH} presents the magnetization vs magnetic field data between 2 and 18\,K. The metamagnetic transitions discussed throughout the paper are evident as kinks and jumps in the plots. Observe how the transitions shift to lower temperatures as temperature increases. Plots for both increasing and decreasing field are displayed each temperature but hysteresis is only observed for low temperature curves and at the III-IV transition near 0.43\,T at 9\,K.

\section{M\"ossbauer and NRIXS}
\label{sec:Appendix_Mossbauer}

\begin{figure*}
	\includegraphics{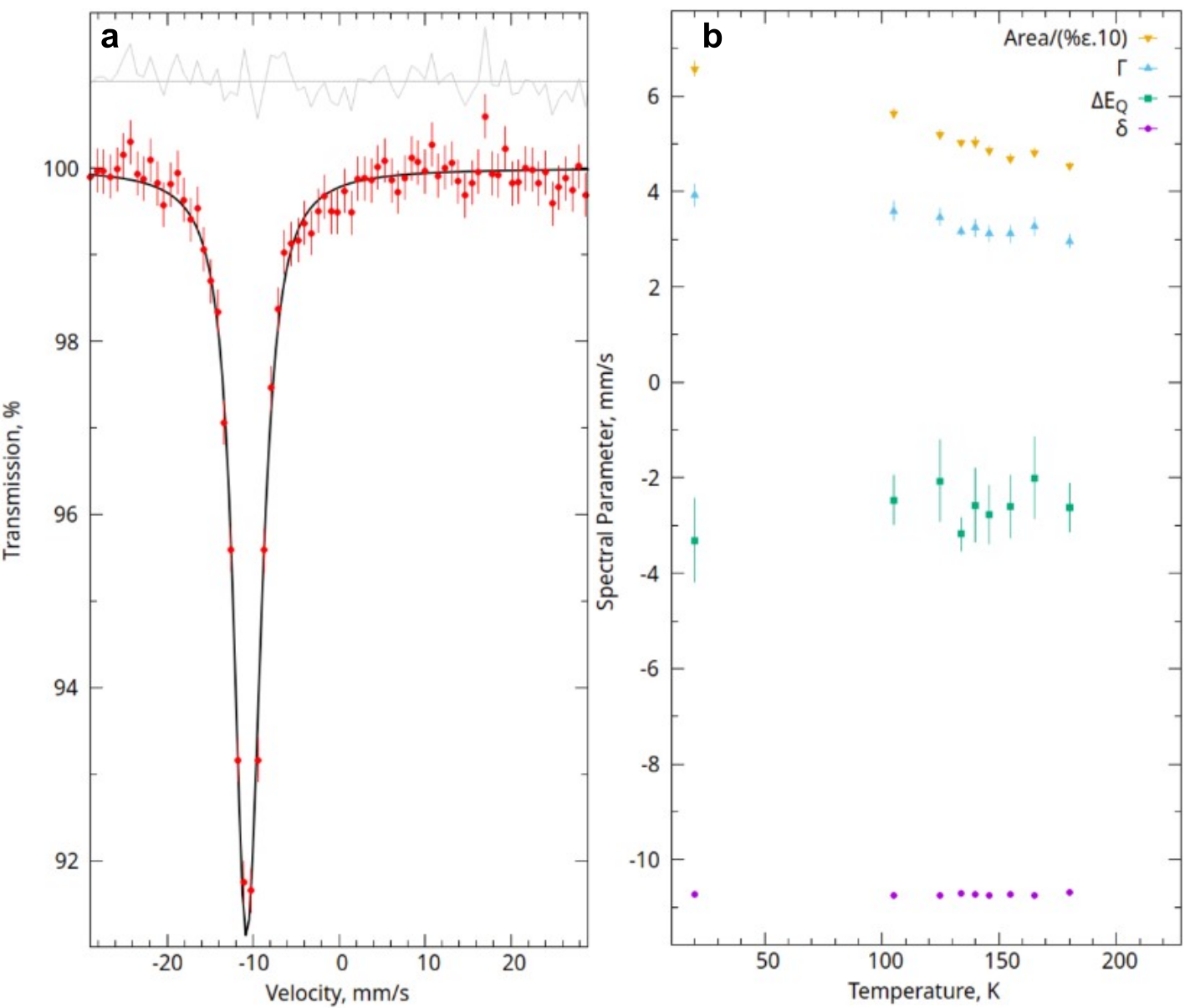}
	\caption{\label{fig:Mossbauer} Temperature dependent $^{151}$Eu M\"ossbauer results. \textbf{(a)} $^{151}$Eu M\"ossbauer spectrum of EuAl$_{4}$ at 146\,K; data in red, fit in black, and residuals in gray. \textbf{(b)} Spectral parameters as function of temperature. $\Gamma$, $\Delta E_{Q}$, and $\delta$ are the linewidth (FWHM), quadrupole splitting, and isomer shift, respectively. The spectral area is in \%\,effect\,mm/s (scaled by a factor 10).
	}
\end{figure*}

M\"ossbauer spectra feature a single absorption line depicted in Fig.~\ref{fig:Mossbauer}\textbf{(a)}. The temperature dependence of the fit parameters appear in Fig.~\ref{fig:Mossbauer}\textbf{(b)}. Detailed analysis reveals broadening by a small quadrupole splitting, which consistently is about -3\,mm/s and an isomer shift of -10.8\,mm/s. The isomer shift is consistent with an earlier report\cite{deVries1983_151EuIsomerShiftsIntermetallics}, and close to a report\cite{Wickman1966_MossbauerHF+IS-MagOrderedEuCompounds} where the reference used for the isomer shift was not specified. Upon cooling, the line-width is increasing due to thickness broadening. 

\begin{figure*}
	\includegraphics{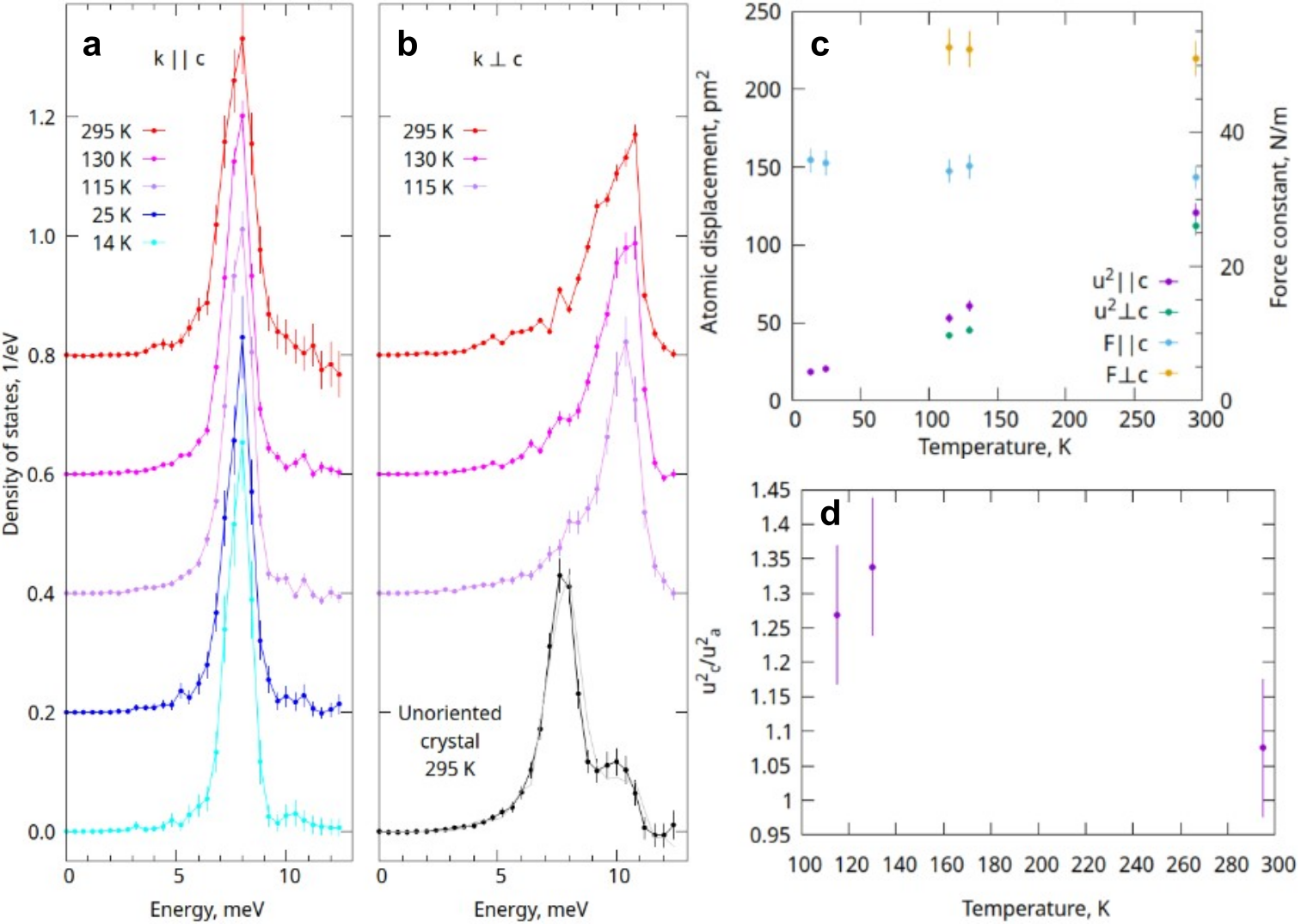}
	\caption{\label{fig:NRIXS} Temperature dependent $^{151}$Eu nuclear resonant inelastic x-ray scattering (NRIXS) results. \textbf{(a)} and \textbf{(b)} density of europium vibrational states parallel and perpendicular to $c$ at the indicated temperatures. The un-oriented crystal data at ambient conditions can be modeled by linear combination (in gray) of data at 295\,K parallel (78\%) and perpendicular (22\%) to $c$. \textbf{(c)} temperature dependence of the Eu force constant, $F$, and dynamic atomic displacement parameter, $u^{2}$, parallel and perpendicular to $c$. \textbf{(d)} the anisotropy in $u^{2}$. A 5\% error bar on $u^{2}$ and $F$ is estimated.
	}
\end{figure*}

The Eu-specific nuclear inelastic scattering for EuAl$_{4}$ was obtained at 115, 130 and 295\,K with the beam parallel and perpendicular to $c$. In addition, data was obtained at 13.5 and 25\,K with the beam parallel to $c$ and one spectrum at 295 K on an un-oriented crystal. The corresponding directionally projected density of Eu vibrational states, $g(E)$, are presented in Figs.~\ref{fig:NRIXS}\textbf{(a)} and \textbf{(b)}. These were obtained by the usual log-Fourier procedure\cite{Chumakov1999_ExperimentalAspectsInelastiNuclearResonanceScattering} using the NISDOS code, a modified version of the program DOS\cite{Kohn2000_DOS,Sergeev_NISDOS}. Anisotropic vibrations for Eu are evidenced by the shape of $g(E)$ which is dominated by a single peak at 8 and 10.5\,meV for the $c$ and $a$ projected data, respectively. 

%

From weighted integrals of the density of vibrational states, we have extracted the atomic displacement parameters and force constants (Fig.~\ref{fig:NRIXS}\textbf{(c)}) using the same procedure as in Ref.~\cite{Moechel2011_LatticeDynamicsYb14MnSb11}. We do not observe any qualitative change in $g(E)$ as a function of temperature, indicating that any change to Eu phonons across the CDW transition is small or limited to a narrow region of reciprocal space. The temperature dependence of the force constant, $F$, reveals the typical softening with increasing temperature. The temperature dependence of the dynamic atomic displacement parameter along $c$ and $a$ indicates the typical increase as function of temperature. Closer inspection reveals that dynamic atomic displacements are mostly isotropic at 295\,K and have 30\% anisotropy below the CDW transition ($u_{c}^{2} \approx 1.3$\,$u_{a}^{2}$) as depicted in Fig.~\ref{fig:NRIXS}\textbf{(d)}. 

\end{document}